\documentclass{aa}  

\usepackage{txfonts}
\usepackage{graphicx}
\usepackage{booktabs}
\usepackage{amsmath}
\usepackage{colortbl}
\usepackage{gensymb}
\usepackage{subfig}
\usepackage{rotating}
\usepackage{amssymb}
\usepackage{latexsym}
\usepackage{ifthen}
\usepackage{enumitem}
\def\gsimeq{\hbox{\raise0.5ex\hbox{$>\lower1.06ex\hbox{$\kern-1.07em{\sim}$}$}}} 
\def\lsimeq{\hbox{\raise0.5ex\hbox{$<\lower1.06ex\hbox{$\kern-1.07em{\sim}$}$}}}

\begin{document} 

  \title{A fast ionised wind in a star-forming quasar system at z$\sim$1.5 resolved through adaptive optics assisted near-infrared data}

%   \subtitle{I. Overviewing the $\kappa$-mechanism}

   \author{       M. Brusa\inst{1,2}\thanks{email:marcella.brusa3@unibo.it}
                    \and 
          M. Perna\inst{1,2}  
          \and
          G. Cresci\inst{3}
          \and
         M. Schramm\inst{4} 
         \and
 I. Delvecchio\inst{5} 
          \and
     G. Lanzuisi\inst{1,2}   
\and
         V. Mainieri\inst{6}
 \and
          M. Mignoli\inst{2}
 \and
 G. Zamorani\inst{2}
\and
      S. Berta\inst{7}
 \and
     A. Bongiorno\inst{8}
          \and
          A. Comastri\inst{2}
\and
 F. Fiore\inst{8} 
 \and
  D. Kakkad\inst{6}
 \and
  A. Marconi\inst{9}
\and
   D. Rosario\inst{10} 
  \and
T. Contini\inst{11,12}
 \and
 F. Lamareille\inst{11}
          }

   \institute{Dipartimento di Fisica e Astronomia, Alma Mater Studiorum, Universit\`a di Bologna,
  viale Berti Pichat 6/2, 40127 Bologna, Italy 
\and INAF - Osservatorio Astronomico di Bologna, via Ranzani 1, 40127 Bologna, Italy 
\and INAF - Osservatorio Astronomico di Arcetri, Largo Enrico Fermi 5, 50125 Firenze, Italy 
\and Frequency Measurement Group, National Institute of Advanced Industrial Science and Technology
(AIST) Tsukuba-central 3-1, Umezono 1-1-1, Tsukuba, Ibaraki 305-8563, Japan
\and Department of Physics, University of Zagreb, Bijenic\v{c}ka cesta 32, HR-10000 Zagreb, Croatia
\and  European Southern Observatory, Karl-Schwarzschild-str. 2,  85748 Garching bei M\"unchen, Germany 
\and Max Planck Institut f\"ur Extraterrestrische Physik, Giessenbachstrasse 1, 85748 Garching bei M\"unchen, Germany
\and INAF - Osservatorio Astronomico di Roma, via Frascati 33,   00078 Monte Porzio Catone (RM) Italy 
  \and Dipartimento di Astronomia e Scienza dello Spazio, Universit\`a degli Studi di Firenze, Largo E. Fermi 2, 50125 Firenze, Italy
\and Centre for Extragalactic Astronomy, Department of Physics, Durham University, South Road, Durham, DH1 3LE, U.K.
\and IRAP, Institut de Recherche en Astrophysique et Planétologie, CNRS, 14, avenue Edouard Belin, F-31400 Toulouse, France
\and Universit\'e de Toulouse, UPS-OMP, Toulouse, France   
}

   \date{Received }

% \abstract{}{}{}{}{} 
% 5 {} token are mandatory

  \abstract
  % context heading (optional)
  % {} leave it empty if necessary  
  {}
  % aims heading (mandatory)
   {Outflow  winds  are  invoked  in  co-evolutionary  models  to  link  the  growth  of  SMBH  and  galaxies  through feedback  phenomena, and from the analysis of both galaxies  and active galactic nuclei (AGN)  samples at z$\sim1-3$, it is becoming clear that powerful outflows may be very common in AGN hosts. High-resolution and high S/N observations are needed to uncover the physical properties of the wind through kinematics analysis.}
  % methods heading (mandatory)
   {We exploited VLT/VIMOS, VLT/SINFONI, and Subaru/IRCS adaptive optics (AO) data to study the kinematics properties on the scale of the host galaxy of XID5395; this galaxy is a luminous, X-ray obscured starburst/quasar (SB-QSO) merging system at z$\sim1.5,$ detected in the XMM-COSMOS field,  associated with an extreme [O II] emitter (with equivalent width, EW, $\sim200$ \AA).
 For the first time, we mapped   the kinematics of the  [O III] and H$\alpha$ line complexes and linked them with the [O II] emission at high resolution. The high spatial resolution achieved allowed us to resolve all the components of the SB-QSO system. }
  % results heading (mandatory)
   {Our analysis, with a resolution of few kpc, reveals complexities and asymmetries in and around the nucleus of XID5395. The velocity field measured via  non-parametric analysis reveals different kinematic components with maximum blueshifted and redshifted velocities up to $\gsimeq1300$ km s$^{-1}$ that are not spatially coincident with the nuclear core.  These extreme values of the observed velocities and spatial location can be explained by the presence of fast moving material.
We also spectroscopically confirm the presence of a merging system at the same redshift as the AGN host.
}
  % conclusions heading (optional), leave it empty if necessary 
   {We propose that EW as large as $>150$ \AA\ in X-ray selected AGN may be an efficient criterion to isolate objects  associated with the short, transition phase of ``feedback" in the AGN-galaxy co-evolutionary path. This co-evolutionary path subsequently evolves into an unobscured QSO, as suggested from the different observational evidence (e.g. merger, compact radio emission, and outflow) we accumulated for XID5395.}


\titlerunning{Feedback in action in XID5395}

   \maketitle
%
%________________________________________________________________

\section{Introduction}

Many of the most successful models of galaxy formation (e.g. \citealt{Hopkins2006a,King2005}) require the presence of  winds, extending over galaxy scales (i.e. $\sim1-10$ kpc and beyond), that are sufficiently fast and energetic  to reproduce the properties observed in the local Universe, such as galaxy colours and counts, and the scaling relations between host galaxies and black hole (BH) properties (e.g. \citealt{Magorrian1998,Marconi2003,Kormendy2013}). These massive gas outflows can indeed be powered by AGN through a radiatively driven process associated with a luminous, obscured, and dust-enshrouded BH accretion phase 
(see e.g. \citealt{Hopkins2008,Fabian2012,King2015}). Alternatively,  massive gas outflows could also be powered by star formation (SF) activity (e.g. \citealt{Heckman1990,Veilleux2005,Geach2014}). % 
Whatever their nature, gas flows are believed to play a pivotal role in shaping galaxies, and much effort, both observational and theoretical, is underway to quantify the relative dominance of the different mechanisms involved.

Galaxy-wide winds at kiloparsec scales \footnote{Winds at much smaller scales, e.g. sub-parsec coming directly from the accretion disk region are also observed in AGN via prominent absorption features in highly ionised lines, e.g. MgII or CIV broad absorption lines quasars (BAL QSOs).} 
in AGN and QSOs hosts have been historically detected %in the ionized gas component 
via observations of broad, blueshifted, and  asymmetric profiles in ionised permitted and forbidden emission lines (e.g. [O III]$\lambda$5007\AA; \citealt{Zamanov2002,Aoki2005}; but see \citealt{Gaskell2013},2015 for a completely different view). These winds are now  also routinely observed in other gas phases (molecular and neutral) with optical/FIR/mm facilities both in local galaxies (e.g. \citealt{Feruglio2010,Rupke2011,Sturm2011,Cicone2014}) and at high-z (e.g. \citealt{Alexander2010,Maiolino2012,CanoDiaz2012,Harrison2012,Genzel2014}).  
 Given that  extreme [O III] line widths such those observed in  local Seyferts and QSOs (FWHM$>$700-1000 km s$^{-1}$; i.e. \citealt{Zhang2011,Villar2011a,Zaurin2013,Harrison2014}) can hardly be ascribed to gravitational motions, these line widths should trace the kinematics of  large-scale energetic gas flows. In these gas flows, where the blue asymmetry is interpreted as fast material moving towards our line of sight, while the receding gas expected in the case of a biconical outflow is preferentially obscured by the host galaxy and generally not observed (but see \citealt{Bae2014,Perna2015}). 

The key ingredient to  reveal unambiguously the presence of AGN-driven winds in systems at the prsk epoch of galaxy evolution (z$\sim1-3$) is spatially resolved NIR spectroscopy, which is capable of directly measuring the existence and extension of an outflow of the warm/ionised gas phase traced by the rest-frame [O III]5007 lines. So far, the most convincing detections of outflows on galactic wide ($\sim5-15$ kpc) scales in radio-quiet AGN in this redshift range have been obtained with seeing-limited IFU data on $\sim$ a dozen unobscured QSOs and/or SMGs (\citealt{CanoDiaz2012,Harrison2012,Cresci2015,Carniani2015,Perna2015_miro}; see also \citealt{Genzel2014}).
%
%                                     Two column figure (place early!)
%______________________________________________ Gamma_1 (lg rho, lg e)
   \begin{figure}
   \centering
  \includegraphics[width=9cm]{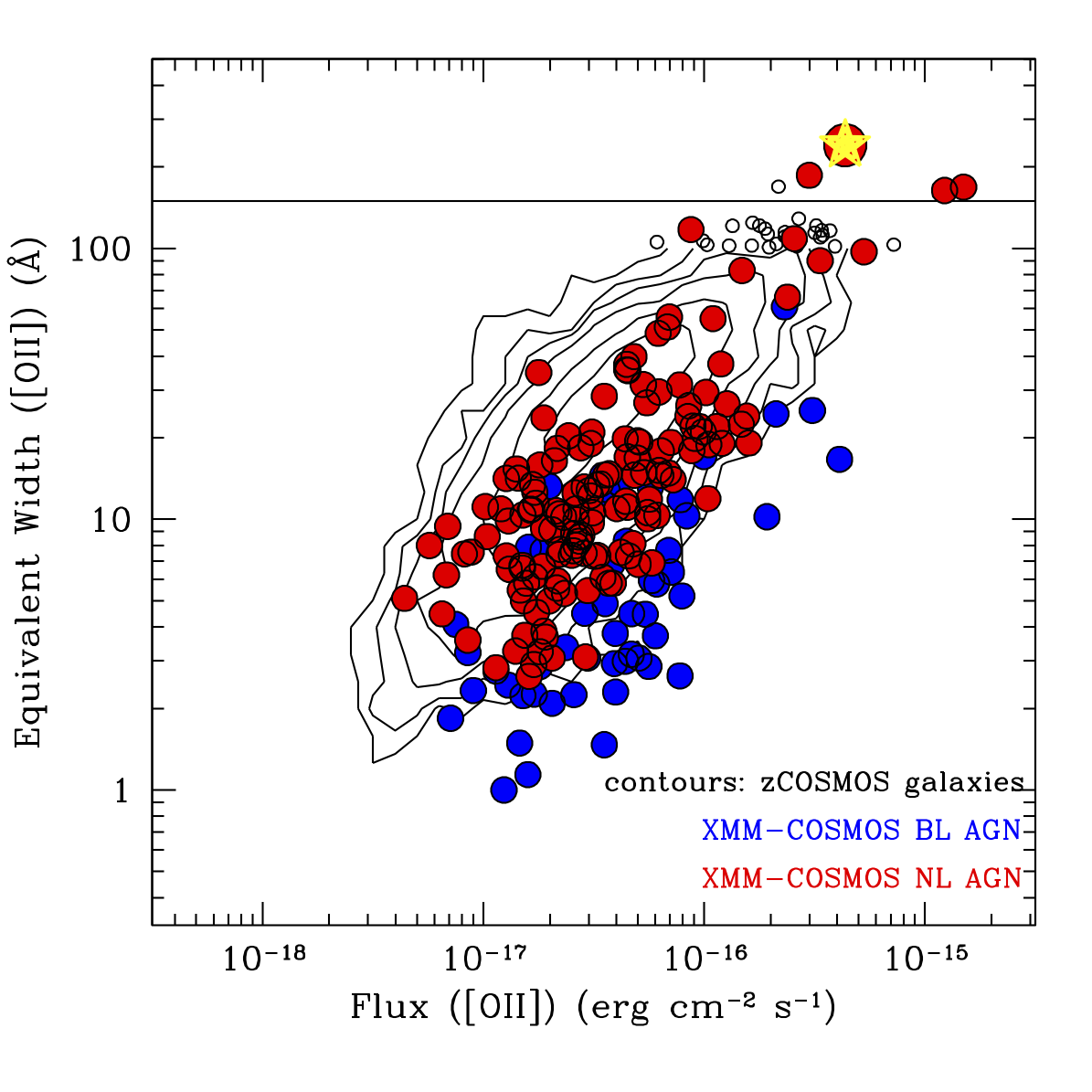}
   \caption{
Rest-frame EW of the [O II] line plotted against the total line flux for S/N$>3$ objects in the zCOSMOS sample within z=0.5-1.5 (7338 objects; black contours and small open circles out of the contours) with  the XMM-COSMOS sources superimposed (blue = 57 Type 1 AGN, red = 152 Type 2 AGN).  XID5395 is highlighted as a yellow star. Out of the five objects with [O II] EW larger than 150 \AA, four are Type 2 QSOs (with L$_{\rm X}\gsimeq10^{44}$ erg s$^{-1}$). XID5395 is the only one with rest-frame EW larger than 200$\AA$ in the entire sample (yellow starred symbol). } 
              \label{selection}%
    \end{figure}
%

%                                     Two column figure (place early!)
%______________________________________________ Gamma_1 (lg rho, lg e)
   \begin{figure*}
   \centering
  \includegraphics[width=18cm]{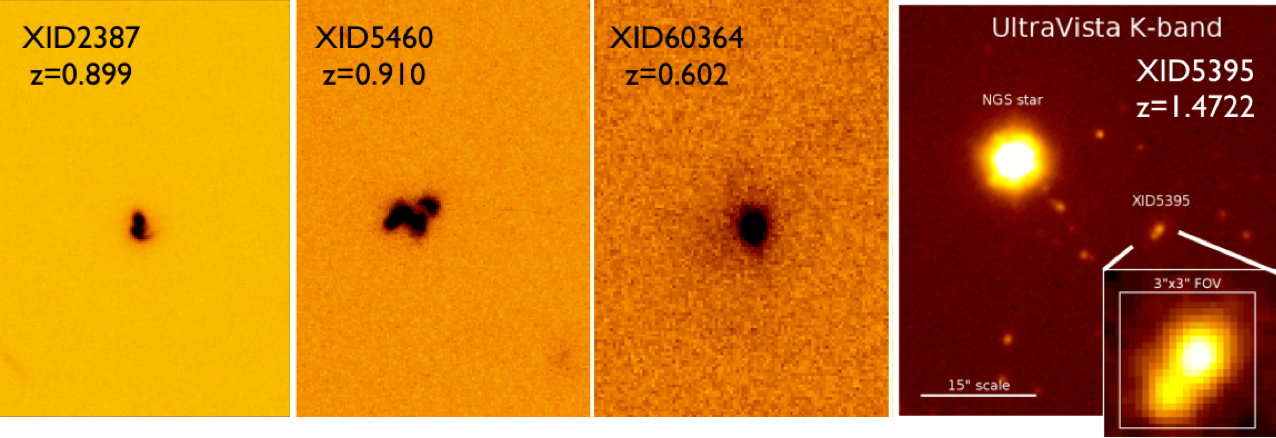}
   \caption{HST/ACS I-band images of the three sources with rest-frame EW$>150$\AA, associated with Type 2 AGN, within the HST COSMOS survey. The XID from the XMM-COSMOS sample and the measured redshfit for each source is plotted at the top of each panel. The last panel shows the Ultravista K-band image of the field around XID5395. The star used for NGS assisted AO observations is also labelled (distance$\sim20$"). The inset shows a zoom on the target with  the 3"$\times 3"$ SINFONI FOV superimposed.}
              \label{ACS}%
    \end{figure*}
Given that  the critical coalescence/blow-out phase is expected to be very short ($<<500$ Myr; see e.g. \citealt{Aversa2015,Volonteri2015}) and X-ray active (e.g. accretion on the BH is at its maximum), 
large area X-ray surveys with the associated high-quality, multi-wavelength data provide the necessary tools to select these very rare sources (see \citealt{Brandt2015} and \citealt{Merloni2016} for recent reviews). If our current understanding of galaxy AGN co-evolution is correct, the X--ray luminous AGN population at z=1-3 should comprise a mixed bag of objects with null to vigorous star formation, depending on the exact phase in which they are caught, the relative intensity of the star formation rate (SFR) and AGN activity, and the amount of obscuration;  a fraction of these X-ray luminous AGN should experience an outflowing wind (see \citealt{King2011,Fabian2012}).

The status of the XMM-COSMOS survey \citep{Hasinger2007,Cappelluti2009,Brusa2010}  in terms of identification, redshift information, and  multi-wavelength classifications is such that it is possible to characterise the SFRs, masses (host galaxy and BH), structure parameters, morphology, and accretion properties of the AGN population with unprecedented detail up to z$\sim$3 (e.g. \citealt{Brusa2010,Bongiorno2012,Lusso2012,Elvis2012,Hao2014}). 
Building on this huge investment of telescopes time, we  pursued a detailed characterisation of the most luminous (L$_X>10^{44}$ erg s$^{-1}$) QSO population missed in pencil beam deep surveys (e.g. \citealt{Mainieri2011,Merloni2014}). 
In particular, the analysis of NIR spectra obtained through an X-shooter campaign of XMM-COSMOS targets selected as outliers in colour-colour diagrams (following \citealt{Brusa2010}) indeed reveals the presence of broad, and blue- and redshifted components, in the [O III] lines ($|\Delta \rm{v}|\sim350-550$ km/s) in six out of eight sources \citep{Brusa2015}. The existence of a kinematically perturbed, ionised component extended over the full galaxy scale in one of these targets (XID2028) has  subsequently been confirmed by slit-resolved spectroscopy \citep{Perna2015} and SINFONI IFU observations \citep{Cresci2015}.

In this paper, we further explore the luminous X--ray population in XMM-COSMOS in conjunction with the zCOSMOS spectroscopic database (\citealt{Lilly2009}; Lamareille et al. in preparation) and we propose a new selection criterion based on the integrated [O II]3727 rest-frame EW to isolate sources with kpc-scale outflows in the obscured SB-QSO phase. We test this criterion on the most extreme object (XID5395) via AO assisted SINFONI and Subaru data. 

The paper is organised as follows: Section 2 presents the target selection, on the basis of a cross-correlation of the X-ray and emission line information and the properties of our test-case AO study. Section 3 presents high-resolution, adaptive optics data obtained with Subaru/IRCS and VLT/SINFONI. Section 4 presents the integrated and resolved flux and kinematics analysis of all  line components, while Section 5 discusses the results in the framework of galaxy-AGN co-evolution. 
We adopt the cosmological parameters $H_0=70$ km s$^{-1}$ Mpc$^{-1}$, $\Omega_m$=0.3 and $\Omega_{\Lambda}$=0.7.
 When referring to magnitudes, we use the AB system, unless otherwise stated.  We assume a \citet{Chabrier2003} initial mass function to derive stellar masses (M$_\star$) and SFRs for the target and the comparison samples. The spatial scale of 1\arcsec\ corresponds to a physical scale $\sim8.5$ kpc at the redshift of XID5395 (z=1.472).

\section{Extreme [O II] emitters and target selection}

High rest-frame EW of oxygen lines (e.g. EW$>$100 \AA)  have been used to isolate objects with extreme levels of star formation  (e.g. \citealt{Amorin2015}; see also  \citealt{Contini2012}). 
We started from the sample of $\sim18000$ spectra observed within the zCOSMOS Bright survey \citep{Lilly2009} and for which we derived accurate line fluxes and parameters from {\tt platefit} measurements, following \citet{Lamareille2009}  and Lamareille et al. (in prep.). We limited the analysis to the $\sim7300$ galaxies within the redshift range 0.5$<$z$<$1.5 with the doublet of [O II]3726,3729 (hereafter; [O II]) measurable, and with signal-to-noise ratio (S/N) on the line larger than 3. 
The values of the rest-frame EW of the [O II] line are plotted against the total line flux for the sample 
in Figure~\ref{selection} (black contours and small open circles out of the contours; from Lamareille et al. in preparation). 
The median redshift is $\langle z \rangle$=0.73, i.e. the zCOSMOS [O II] sample is skewed at z$<1$ and dominated by the large z=0.67 and z=0.73 structures known in the field \citep{Scoville2007_struct}.

We cross-correlated the [O II] emitters catalogue with the XMM-COSMOS identifications catalogue \citep{Brusa2010} and we found 209 matches, corresponding to $\sim3$\% of the overall population.
All the XMM-COSMOS AGN are also plotted in Figure~\ref{selection}: Type 2/narrow line AGN (NL AGN) are indicated with red circles, while Type 1/broad line AGN (BL AGN) are indicated with blue circles.  
When considering objects with line fluxes larger than $10^{-17}$ erg cm$^{-2}$ s$^{-1}$ (at lower fluxes there are basically no AGN in our sample), AGN have lower EWs (EW$_{med}$=10.8\AA) than the overall source population (EW$_{med}$=24.0\AA), and, among the AGN population, BL AGN have considerably lower EW (EW$_{med}$=4.9\AA) than NL AGN (EW$_{med}$=12.5\AA). 
This is expected, given that the nuclear emission contributes to the continuum used to measure the EW in Type 2 AGN, and  dominates in Type 1 AGN.  

The fraction of  X-ray detected [O II] emitters significantly changes when progressing to extreme EWs: among the  32 objects with [O II] EW larger than 100\AA, six are associated with XMM-COSMOS AGN ($\sim$19\%). 
This fraction further increases dramatically when going to EWs$>150$\AA: four out of five objects (80\%) are X--ray AGN and  all of them are classified as QSOs (L$_{\rm X}\gsimeq10^{44}$ erg s$^{-1}$) on the basis of the intrinsic X-ray luminosities. Four out of five sources are also [NeV] emitters, and are part of the sample of  zCOSMOS selected [NeV] emitters whose properties were discussed in Mignoli et al. (2013) and Vignali et al. (2014). 

The HST/ACS cutouts of the three z$<1$ sources with EW$>150$\AA\ and associated with X-ray detected sources are shown in Figure~\ref{ACS} (first three panels) with the corresponding XID from the XMM-COSMOS catalogue and the associated redshifts from the zCOSMOS spectra. All three sources show clear signs of disturbed morphology with tidal features, which is evidence for double nuclei and extended emission.  We therefore propose that high [O II] EW ($\gsimeq150$\AA) can be used as a reliable proxy to catch star-forming galaxies associated with type 2 obscured QSOs, most likely in a merger phase.

%__________________________________________________________________

\subsection{XID5395: a unique SB-QSO system}

The source with largest EW in the entire sample is the XMM-COSMOS source XID5395 (RA=10:02:58.43, DEC=02:10:13.9), a QSO at z=1.472 as measured in the zCOSMOS catalogue (the highest redshift in the [O II] sample) with an intrinsic rest-frame 2-10 keV luminosity of $\sim7.5\times10^{44}$ erg s$^{-1}$ and high obscuration, N$_{\rm H}\sim10^{23}$ cm$^{-2}$ (from the analysis presented in \citealt{Lanzuisi2015}). 
The EW measured from the automatic Platefit procedure \citep{Lamareille2009} is 250\AA\ and is plotted in Figure~\ref{selection} (yellow star). 

Unfortunately, this source lies outside the HST/ACS coverage of the central 1.7 degrees of the COSMOS field \citep{Scoville2007,Koekemoer2007} and therefore lacks high-resolution rest-frame UV imaging (but see also Figure~\ref{cutouts} and Section 4.1). In Figure~\ref{ACS} (right-most panel) we show the highest resolution image available for XID5395 prior to this work (UltraVISTA data release 1, DR1; \citealt{McCracken2012}) of the field around the QSO and with a zoom on a 3"x3" scale (inset).  The UltraVISTA image suggests a complex morphology, possibly associated with two components separated by $\sim1"$ ($\sim8.5$ kpc).

%
%                                                One column figure
%----------------------------------------------------------- S_vib
   \begin{figure}
   \centering
  \includegraphics[width=8.3cm]{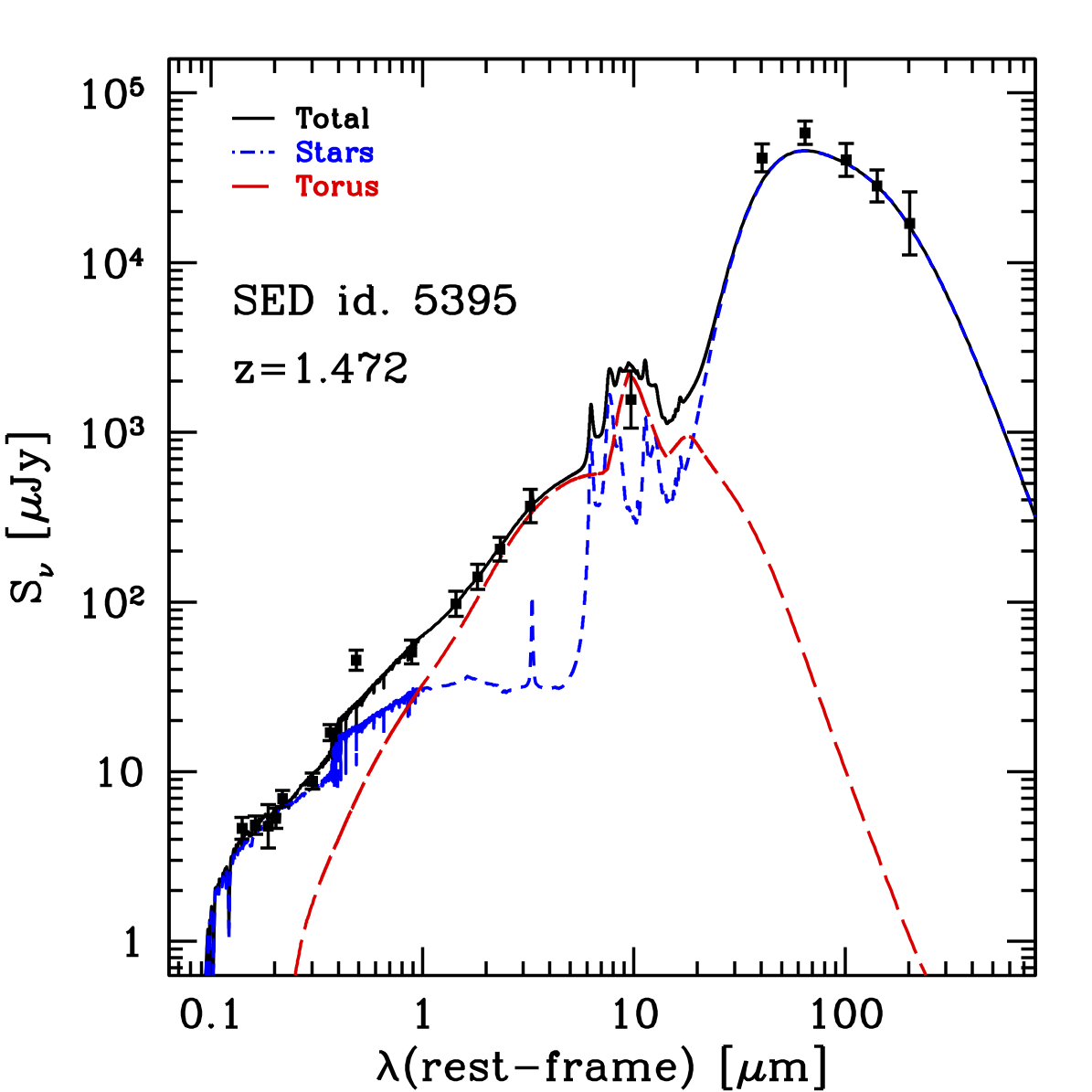}
      \caption{SED decomposition obtained with the code by \citet{Berta2013}. Filled squares represent the available photometric data, long-dashed red, dashed blue, and solid black lines represent the AGN+torus component, the emission by stars and dust associated with star formation, and the sum of the two, respectively.
Bands are, from left to right, u,b,v,g,r,i,z,J,K, IRAC3.6, IRAC4.5, IRAC5.8, IRAC8.0, MIPS24, PACS100, PACS160, SPIRE250,SPIRE350,  SPIRE500.  The  photometric point at $\lambda$(rest)$\sim5000$ \AA\ is contaminated by broad and bright [O III] emission (see Section 4) and, therefore, is above the best-fit model.
    }
         \label{sedfit}
   \end{figure}

XID5395 shares the same level of intrinsic X-ray luminosity as other luminous red QSOs at the same redshift for which outflows in the ionised gas components have been reported (e.g. the X-shooter targets presented in \citealt{Brusa2015}). However, in contrast to these red QSOs hosted in main-sequence (MS) galaxies, XID5395 is heavily obscured in the X-rays (N$_{\rm H}\sim1.2^{+0.9}_{-0.5}\times10^{23}$ cm$^{-2}$; to be compared to $\sim10^{21}-10^{22}$ cm$^{-2}$ reported in \citealt{Perna2015} for the \citealt{Brusa2015} sample) and it is hosted in a starburst galaxy (i.e. MS outlier at z=1.5).  
We fitted the broadband SED of XID5395, extending from UV to FIR and taking advantage of Herschel PACS and SPIRE data, using the code by \citet{Berta2013}. Originally inspired by \citet{Dacunha2008}, this code combines the emission from stars, dust heated by star formation, and a possible AGN/torus component. 
The multi-wavelength photometry extracted from the COSMOS database, and the associated SED fitting decomposition is shown in Figure~\ref{sedfit}. 
The FIR bump is very well reproduced by a starburst template with SFR$\sim370$ M$_\odot$ yr$^{-1}$ (see also \citealt{Delvecchio2014}), while the optical and NIR part of the SED returns a stellar mass of M$_\star\sim7.8\times10^{10}$ M$_\odot$. The AGN emission dominates in the Spitzer bands.
We further discuss the host galaxies properties of XID5395 in Section 5.4.

Finally, XID5395 is a bright radio source with a 1.4GHz flux of 2.26$\pm0.057$ mJy (from the VLA-COSMOS survey; \citealt{Schinnerer2010}), corresponding to a radio power P$_{1.4 GHz}\sim3\times10^{25}$ W Hz$^{-1}$. Previous work \citep{Chiaberge2009} classifies XID5395 as `low power FRI radio source' with a compact morphology and without any evidence of the presence of radio jets (their source \#37). 
The source is also detected in the VLA-COSMOS 3GHz Large Project (Smol\v{c}i\'c et al., in preparation) and in the high-resolution VLBA observations (Herrera-Ruiz et al., in prep): from the ratio of total-to-peak flux the radio emission is resolved over the VLA beam (0.7\arcsec), but the presence of a faint compact core is confirmed (see Delvecchio et al. in preparation).
On the basis of the radio-to-infrared flux ratios (q$_{24}\sim0$ with a 24 $\mu$m flux of 1.58 mJy as measured by MIPS; q$_{100}\sim1.24$ with a measured 100 $\mu$m flux from PACS of 39.6 mJy) XID5395 is classified as `radio-active' (see \citealt{Magliocchetti2014}). The radio power exceeds  that expected from the \citet{Bell2003}  relation given the observed FIR luminosity by one order of magnitude, and this is most likely due to the AGN activity. 

\subsubsection{Non-parametric analysis of the [O II] line}
The left panel of Figure~\ref{oxygen} shows a zoom of the optical VIMOS spectrum from the zCOSMOS survey, which is centred around the [O II] emission line.  In addition to the unusually bright component in the [O II] emission, XID5395 also shows a  broad and asymmetric profile of the line. 

We used non-parametric measurements of the [O II] emission line profile following \citet[; see also \citealt{Rupke2013,Zakamska2014}]{Perna2015}. 
Briefly, we measured the velocity $v$ at which a given fraction of the total (best-fit) line flux is collected, using a cumulative function $F(v)=\int_{-\infty}^{v} F_v(v')\,dv'$.
The  best-fit profile and the position of $v$=0 of the cumulative flux are obtained from a multicomponent fit of the 1\arcsec integrated spectrum, and adopting the systemic redshift of z=1.4718 (see Section 4.1)\footnote{We use a single Gaussian line to model the doublet of the [O II]3727  emission, which has an intrinsic separation of $\sim$250 km/s with a relative ratio that depends on the density of the gas. The natural profile is not Gaussian. However, we note that the intrinsic doublet separation does not change the non-parametric measurements too much: Indeed, assuming a narrow component of 500 km/s per each of the emission lines in the doublet, the total FWHM measured from a single Gaussian component would be $\sim600$ km s$^{-1}$, which is much lower than the non-parametric measurements discussed below. See also the discussion in Zakamska \& Greene (2014).}. In this way, the velocity measurements are less sensible to the S/N of the data at the observed wavelengths, but are dependent on the goodness of the overall fit profile.
We carried out the following velocity measurements on the best-fit [O II] line profile:

%                                     Two column figure (place early!)
%______________________________________________ Gamma_1 (lg rho, lg e)
   \begin{figure*}
   \centering
  \includegraphics[width=17cm,angle=0]{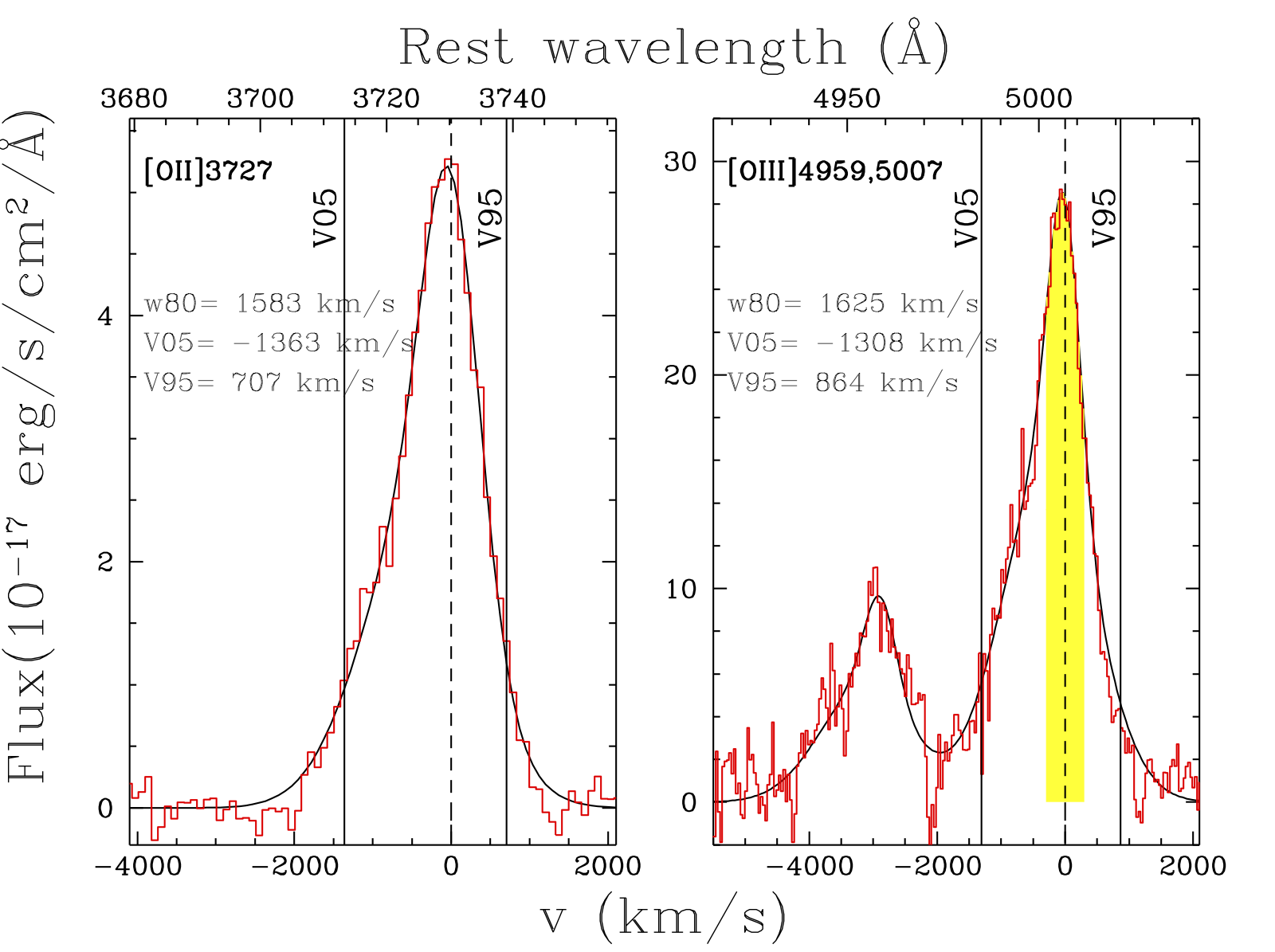}
   \caption{{\it Left panel}: non-parametric analysis of the [O II]3727 doublet emission-line profile for XID5395 (red histogram) as observed in the VIMOS spectrum. The solid curve is the fit to the profile obtained with a combination of two Gaussian lines (not shown here; see also Figure~\ref{spectrum_sim}). The vertical solid lines show different percentiles to the flux contained in the overall emission-line profile (v$_{05}$ and v$_{95}$) as labelled. The long-dashed line indicates the v=0 systemic as determined in Section 4.1.  {\it Right panel}: same as the left panel, but around the [O III]4959,5007 emission lines, from the SINFONI spectrum, extracted from an aperture of 1\arcsec  diameter (see Section 4.1 for details). A clear asymmetry is present in both the [O II] and [O III] emission lines with similar kinematics parameters (as labelled). The yellow shaded region highlights the wavelengths interval corresponding to the peak of the [O III] emission (+/- 300 km/s) in which the map shown in Figure~\ref{lines_syst} (left) is integrated. 
}
              \label{oxygen}%
    \end{figure*}

\begin{enumerate}
\item w80: The line width comprising 80\% of the flux, that for a Gaussian profile is $\sim10$\% larger than the FWHM value.  w80 is defined as the difference between the velocity at 90\% (v$_{90}$) and 10\% (v$_{10}$) of the cumulative flux, respectively;
\item V$_{\rm max, blue}$(=v$_{05}$): The maximum blueshift velocity parameter defined as the velocity at 5\% of the cumulative flux; and
\item V$_{\rm max, red}$(=v$_{95}$): The maximum redshift velocity parameter defined as the velocity at 95\% of the cumulative flux.
\end{enumerate}

These non-parametric values are shown in the left panel of Figure~\ref{oxygen}. We clearly recover the broad profile with w$_{80}\sim1600$ km s$^{-1}$ and extending to absolute velocities higher than 1000 km/s both in the blue and red tails. 
When compared to the values obtained in a sample of luminous obscured QSOs from the SDSS survey presented in \citet{Zakamska2014}, XID5395  qualifies  as a clear outlier;
both in the zCOSMOS (galaxy and AGN) and in the SDSS (luminous obscured QSOs) samples there are no other sources with the same extreme properties of the [O II] line.

%______________________________________________________________

\section{High-resolution adaptive optics data} 
The pieces of evidence accumulated so far, such as the blueshifted wing in the [O II] emission, simultaneous BH accretion and star formation, and possible merger status, and most importantly the presence of a bright star at a suitable distance needed for AO-assisted observations (see Figure~\ref{ACS}, right panel),  make XID5395 the perfect target for a detailed study of feedback from AGN in the transition phase of galaxy-AGN coevolution. 
We therefore mapped the AGN and the host galaxy in XID5395 both in imaging and spectroscopy using SUBARU IRCS and SINFONI in AO assisted mode.

\subsection{SUBARU IRCS imaging}
 
We used the Infrared Camera and Spectrograph (IRCS; \citealt{Kobayashi2000}), installed at the Infrared Nasmyth focus of the Subaru Telescope, in combination with its adaptive optics system AO188 \citep{Hayano2010}, to obtain high-resolution infrared images of the field around XID5395. 

We obtained the observations  on December 15, 2014.  We chose a resolution of 52mas/pixel for a corresponding 54\arcsec field of view. 
We performed observations  in imaging mode with IRCS equipped with the H-band (1.33-1.93 $\mu$m range) and K-band (1.86-2.54 $\mu$m range) broadband filters.
The H-band data contain the H$\alpha$ emission (redshifted at 1.622 $\mu$m), and can be directly compared to the SINFONI H-band data (see below). The K-band data instead  should be free from bright emission lines and should genuinely sample the  stellar population of the host galaxy. 

We reduced the data using a modified version of the IRAF data reduction pipeline for SUBARU/IRCS. The reduction  includes the creation of a bad pixel mask and a skyflat from the data themselves, the flat fielding, sky subtraction, estimate of the dither offsets, and  final stacking of the data. 

Conditions during the observations were variable with a 0.7-1.0\arcsec\ natural seeing.
In the K band, all observations were performed using the Laser Guide Star (LGS) mode. We used  a 9 point dither pattern (with 5\arcsec\ stepsize) and 20s exposure time each and a bright star (11 mag in H band) separated 20\arcsec for tip-tilt correction. In  total, we used 45 frames in the final stacking for a total exposure of 15 minutes. The best image resolution at the TTS position is $\sim$0.1\arcsec\ after AO correction. The resolution at the target position is about $\sim$0.2\arcsec. 
In the H band we also adopted a 9 point dither pattern with 50s exposure time. We used 27 (good) frames in the final stack for a total exposure of 22.5 minutes. We achieve a slightly worse resolution ($\sim$0.24\arcsec) in the H band with respect to the K band.

%                                                One column figure
%----------------------------------------------------------- S_vib
   \begin{figure*}
   \centering
  \includegraphics[width=18.3cm]{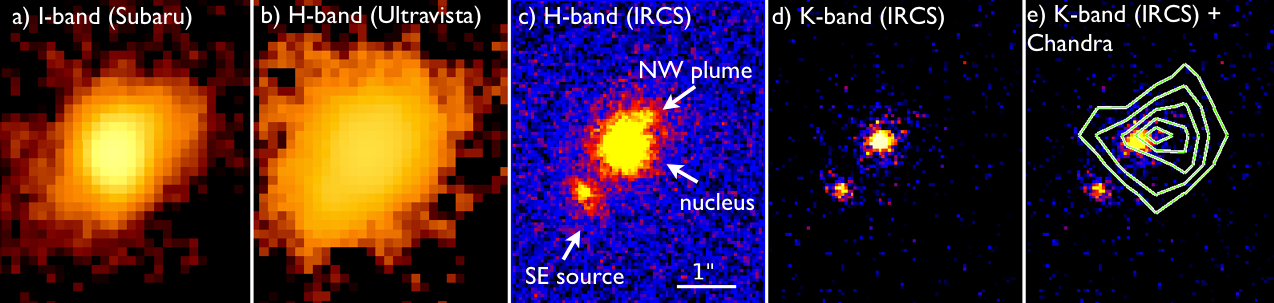}
      \caption{From left to right, image cutouts of XID5395 at increasing wavelengths and resolution: a) SUBARU Suprimecam I band (FWHM$\sim$1\arcsec); b) UltraVISTA H band (FWHM$\sim$0.7\arcsec); c) SUBARU IRCS H band (FWHM$\sim$0.2\arcsec); d)  SUBARU IRCS K band (FWHM$\sim$01\arcsec); e) same as panel d) with  the X--ray contours superimposed (from {\it Chandra} Legacy data; \citealt{Civano2016}). The cutouts have a size of $\sim4$\arcsec across (e.g. comparable to the SINFONI FoV).  The 1\arcsec scale is given for reference in panel c). In the same panel we also indicate different components discussed in the text, as labelled. North is up and east is left. Chandra contours are not corrected for astrometry.
    }
         \label{cutouts}
   \end{figure*}

The images were brought to the COSMOS reference system with an astrometric correction based on UltraVISTA data through a match between the coordinates of a bright point-like source detected in all the images at a projected distance of 10.5$\arcsec$ from the QSO. The overall astrometric accuracy of our imaging is  $\sim$0.1\arcsec.
The IRCS H-band and K-band images are shown in Figure~\ref{cutouts} (panel c and d). They are compared to the Subaru Suprimecam I band data (panel a; resolution/FWHM$\sim$1\arcsec) and the UltraVISTA H-band data (panel b; resolution/FWHM$\sim0.7$\arcsec). 

The Subaru IRCS data at higher resolution clearly resolve two objects unresolved in the UltraVISTA image. These two objects are labelled as ``nucleus" and ``south-east (SE) source" in Figure~\ref{cutouts}, and they are separated by $\sim0.6\arcsec$. In addition, an extended feature can be seen in the H-band high-resolution IRCS data (third panel in Figure~\ref{cutouts}), extending from the nucleus towards the north-west (NW) direction. We name this feature ``NW plume". Instead, the IRCS K-band map, which is free of contamination from bright emission lines, appears much more compact ($<0.4-0.5$\arcsec) and symmetric, although the overall S/N is worse compared to the H band.
The coordinates of the centroids and the magnitudes in the H and K band of all these three components are reported in Table~1. The nucleus clearly dominates the photometry of the entire system, and the value obtained for the nucleus is consistent with the total magnitudes measured in UKIRT and  CFHT observations (H=19.7,K=19.6; see also Laigle et al., in preparation for UltraVISTA data). 

In the last panel of Figure~\ref{cutouts}, we plot again the K-band IRCS data, superimposed with the {\it Chandra} Legacy X-ray contours (full band; taken from \citealt{Civano2016}). The high-resolution {\it Chandra} image also unambiguously associates the observed X--ray emission with the  brightest source detected in our deep Subaru imaging. 

\subsection{SINFONI observations and spectral analysis}

The observations were performed in AO-assisted mode using the near-infrared (NIR) Integral Field Unit (IFU) Spectrometer SINFONI of the Very Large Telescope (VLT), during the period 94 (from 2015-02-10 to 2015-03-28). 
XID5395 was observed in two filters, J and H, to sample the [O III]+H${\beta}$ complex (redshifted in the J band, at $\sim$1.22 $\mu$m) and the H${\alpha}$+[N II] complex (redshifted in the H band, at $\sim$1.69 $\mu$m). We collected a total of six (2) sky and 12 (4) on-source exposures of 600 s each for the J band (H band). The total integration times on source were 120 minutes for the J-band data,  and 40 minutes for the H band. 

We used a field of view (FoV) of 3x3\arcsec in a 2D 64x64 spaxel frame. 
The spectral resolutions are R$\sim$1800 for J and R$\sim$2500 for H. 
We achieved a spatial resolution of 0.33 and 0.45\arcsec for the J and H band, respectively, based on the  point spread functions (PSF) obtained in AO-mode taken just before the QSO observations. This roughly corresponds to 2.8 and 3.8 kpc at the redshift of the XID5395, z=1.5. 
Standard stars and the respective sky frames were also observed to flux-calibrate the data. 

The data reduction process was performed using ESOREX (version 2.0.5).
We used the ESOREX jitter recipe to reconstruct the data cube and the 
IDL routine "skysub'' \citep{Davies2007} to remove the background sky emission. Then, we used our own IDL routines to perform the flux calibration on every single cube and to reconstruct a final datacube for each object, co-adding the different pointings. 
To  establish an absolute flux calibration, we matched synthetic J- and H-band photometry obtained from SINFONI 1\arcsec integrated spectra with COSMOS J- and H-band photometry. We checked the wavelength calibration against known sky lines in the IR and found good consistency. Finally, we applied an astrometric calibration of the J and H data cube using the IRCS H band.

We detected no continuum emission overall the FoV in either SINFONI band. We measure a 3$\sigma$ upper limit of 0.6 (0.9) $\times10^{-17}$ erg s$^{-1}$ cm$^{-2}$ $\AA^{-1}$  in J (H) band at the position of the nucleus.

%
%                                                One column figure
%----------------------------------------------------------- S_vib
   \begin{figure*}[!t]
   \centering
  \includegraphics[width=18.cm]{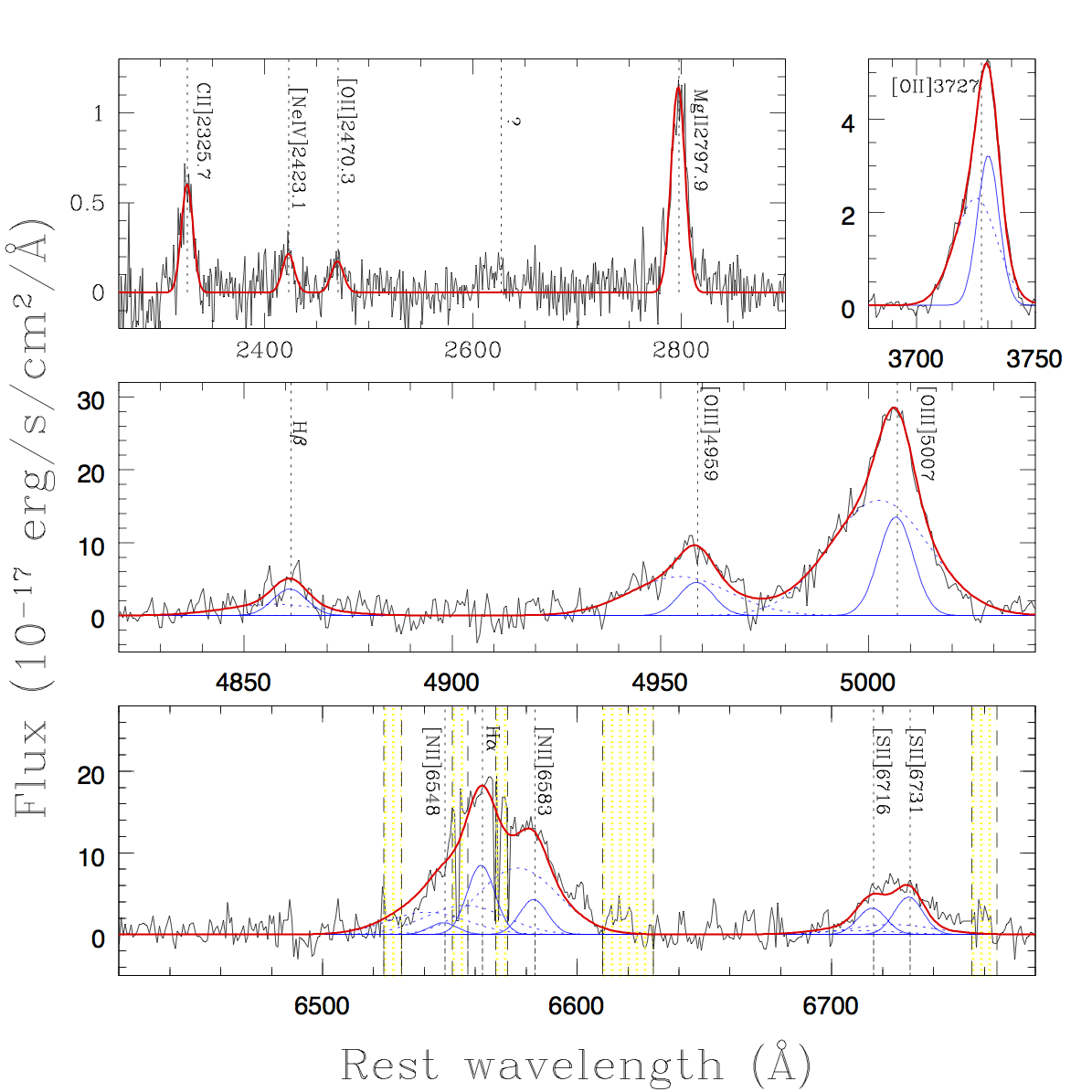}
      \caption{{\it Top:} VIMOS spectrum shows in two zoomed regions the 2300-2900 rest-frame wavelength range and the [O II] emission. {\it Central:}  SINFONI J-band spectrum extracted within an aperture of 1\arcsec\ diameter from the nucleus around the H$\beta$+[O III] emission lines region. {\it Bottom:}  SINFONI H-band spectrum extracted within an aperture of 1\arcsec\ diameter from the nucleus around the H$\alpha$+[NII]+[SII] emission line region.
Spectra are plotted in the rest-frame system; continuum emission has been subtracted from the VIMOS spectrum. 
Superimposed on the spectra are the best-fit components of our simultaneous, multicomponent fit (solid and dashed blue curves). 
The red solid curves represent the sum of all model components. 
From left to right, top to bottom, black dotted lines indicate the wavelengths of the C II], [NeIV], [O II]2471, MgII, [O II]3727, H$\beta$, [O III]4959,5007, [NII]6548, H$\alpha$,[N II]6583, and [S II]6716,6731 emission lines. Yellow shaded regions  indicate wavelength of intense sky lines in the NIR spectra, affecting mostly the H$\alpha$ line, which were excluded from the fit. The emission line at $\lambda\sim2628$ marked with ``?'' is also observed in the near-UV spectrum of  NGC 1068 (see e.g. Raemer, Ruiz, \& Crenshaw 1998) without an associated idenfication.
    }
         \label{spectrum_sim}
   \end{figure*}
%
%______________________________________________________________

\section{Spatial and spectral analysis}
The blueshifted [O II] wing seen in the VIMOS spectrum can be ascribed to the presence of an outflowing wind, which should reveal itself also in the [O III]5007 line  with velocities up to 1500-2000 km s$^{-1}$,  as suggested for example by the analysis of \citet{Zakamska2014}. To confirm the existence of the outflow and further constrain its properties, we  map the [O III]5007 line kinematics in the J-band SINFONI datacube. We also mapped the complex H$\alpha$+[NII] emission sampled by our shallower H-band observations. 

\subsection{Integrated VIMOS and SINFONI spectra}

%
%                                                One column figure
%----------------------------------------------------------- S_vib
   \begin{figure*}
  \includegraphics[width=9.4cm,angle=180]{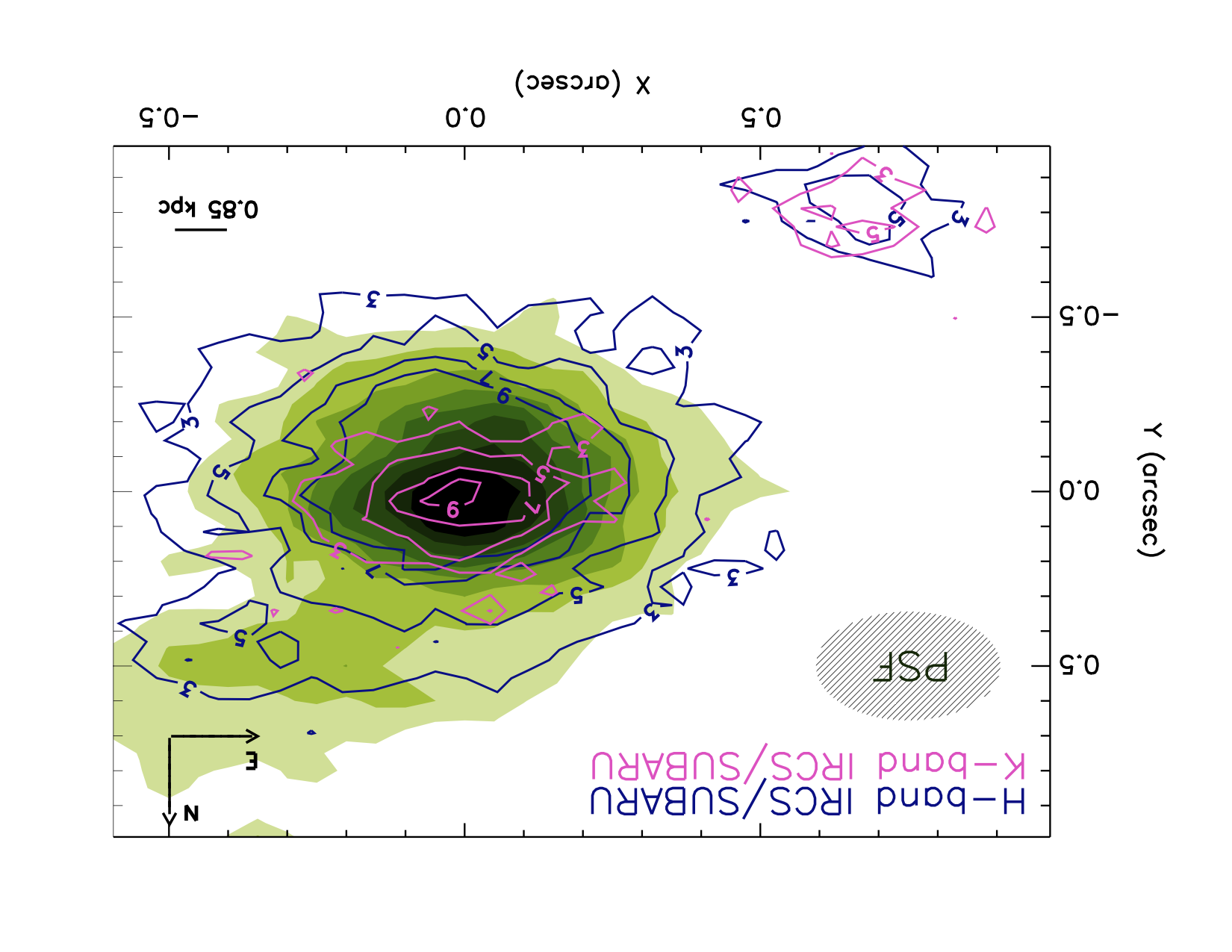}
  \includegraphics[width=9.4cm,angle=180]{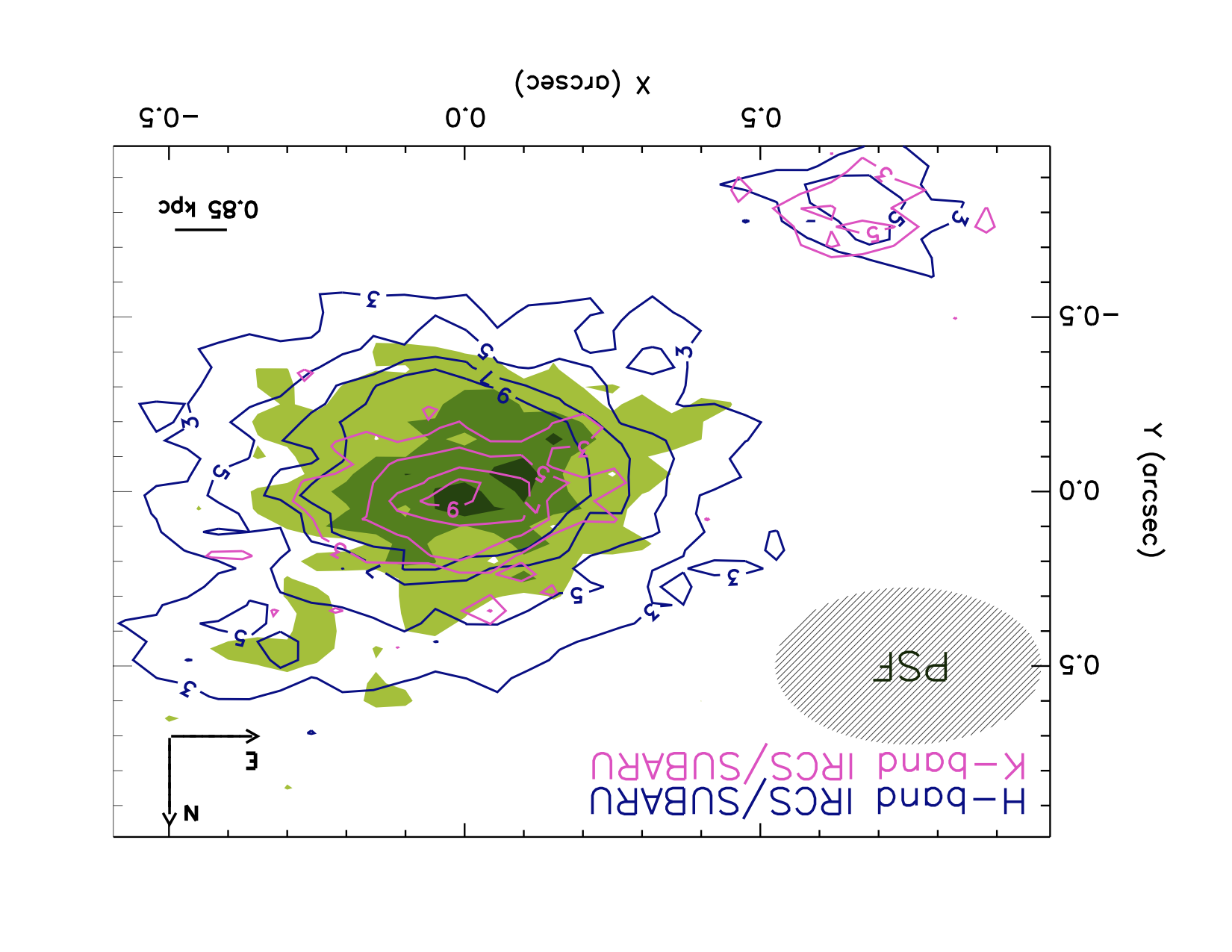}
      \caption{(Left) Contours of the Subaru/IRCS H-band (blue) and K-band (magenta) emission taken from Figure~\ref{cutouts} overlaid on the [O III]5007 channel map obtained integrating the SINFONI datacube on the line core ($\lambda$=1236.1-1238.9 nm; corresponding to the rest-frame range 5001-5012 \AA). All contours and colour level starts from 3$\sigma$ with steps of +2, as labelled.
Spatial scale is given in arcsec. The [O III] systemic emission is overall symmetric, except for a plume extending beyond 0.5\arcsec\ ($\sim4.5$ kpc) in the north-west direction. (Right): same, but for the H$\alpha$ line, collapsing the channels in the wavelength range $\lambda$=1620.6-1624.0 nm (corresponding to the rest-frame range 6556-6570 \AA). The same plume is also seen in the IRCS H-band data (sampling the H$\alpha$ region), but it is not seen in the IRCS K-band data. 
    }
         \label{lines_syst}
   \end{figure*}

From the SINFONI datacubes we extracted a spectrum from an aperture of 1\arcsec  diameter around the nucleus, which is similar to the aperture used to extract the VIMOS spectrum. The full spectrum combined from the J- and H-band SINFONI datacubes, together with the VIMOS spectrum, is shown in Figure~\ref{spectrum_sim}. In the top panel we show a zoom on the spectral region from CII]2326 to MgII (left) and around [O II]3727 (right), in the middle panel around H$\beta$+[O~III], and in the bottom panel on the H$\alpha$+[N II]+[S II] emission lines. Following the prescriptions presented in Brusa et al. (2015; see also \citealt{Perna2015}), we performed  a simultaneous fit with two Gaussian components to reproduce the line profiles. 
We modelled all the emission lines in the entire wavelength range covered by VIMOS, SINFONI J- and H-band spectra with the same kinematic model. 

We used a narrow component (NC; FWHM $\lesssim550$ km s$^{-1}$) to account for the NLR in virial motion, and we also used an outflow component (OC; FWHM $>$550 km/s).
We imposed three constraints in accordance with atomic physics:
1) the flux ratios between [O III]4959,5007, [NII]6548,83 were set at 1:2.99, while the flux ratio between the [SII]6716,6731 was constrained to be within the range 0.44-1.42;
2) the width of a given kinematic component was forced to be equal for all the emission lines;
3) the relative wavelength separations between the emission lines of a given kinematic component were constrained to be equal to the laboratory ratios; the velocity shift $\Delta v$, defined as the shift between the line centroids of the two OC and NC, was allowed to vary.
The [O II]3726,3728 doublet was treated as a single line with 3727.42$\AA$ as the rest-frame wavelength. 
The doublet separation is too small to allow a constraint of both NC and OC components. Therefore, in the simultaneous fit, we leave to vary the width of its NC profile, independently from the other Gaussian profiles of the same kinematic component to take  the blended nature of the doublet into account.

We adopted as systemic redshift that corresponding to the consistent fit of all the narrow Gaussian components in the SINFONI spectra (z=1.4718$\pm$0.0001 where the errors are from MC simulations; see \citealt{Perna2015}); in fact, a possible wavelength shift between the  spectra of two different instruments (VIMOS and SINFONI) could affect  the  rest-frame wavelength. Therefore, in the simultaneous fit the shifts between the NC of each rest-frame optical and rest-frame NIR emission line were constrained independently. We found a VIS-NIR shift of about 240 km/s ($\Delta$z=0.002). 

The results of the emission line fits (e.g. the fluxes in the H$\alpha$, [O III] and [O II] emission lines, and the centroids of the different components) are reported in Table 1. We measure a ratio of the total flux of the [O III] and [O II] lines of the NC, [O III]/[O II]$_{NC}\sim$4, in between  the value measured in the average spectrum of SDSS Type 2 QSOs (\citealt{Zakamska2003}; [O III]/[O II]$\sim2$) and broad line quasars (\citealt{Vandenberk2001}; [O III]/[O II]$\sim6$). Instead, the flux ratio in the OC component is twice as large  ([O III]/[O II])$_{OC}\sim$9. Such extreme values of the flux ratios are more similar to what has been observed in high-z Lyman alpha emitters (LAE) and Lyman-break galaxies (LBG; see \citealt{Nakajima2014}) or green peas galaxies \citep{Cardamone2009}. 

We then applied the same non-parametric analysis described in Section 2.1 to the [O III] lines in the spectrum extracted from the 1\arcsec\ aperture. The same asymmetry present in the [O II]3727 emission line doublet, with similar kinematics parameters, is found. A zoom on the [O III]4959,5007 emission lines in the SINFONI spectrum is shown in the right panel of Figure~\ref{oxygen}, with non-parametric quantities labelled.  

Finally, we used the Balmer decrements F(H$\alpha$)/F(H$\beta$) to estimate the reddening in the overall system.  Assuming the Case B ratio of 3.1 (Gaskell \& Ferland 1984) and the SMC dust-reddening law, we derived a V-band extinction A$_V$=0.8$\pm0.3$, in reasonable agreement with the extinction derived from the SED fitting of the combination of host and galaxy templates (A$_V\sim$1.1, assuming a \citealt{Cardelli1989} extinction law).

   \begin{figure*}
   \centering
  \includegraphics[width=15.5cm,angle=0]{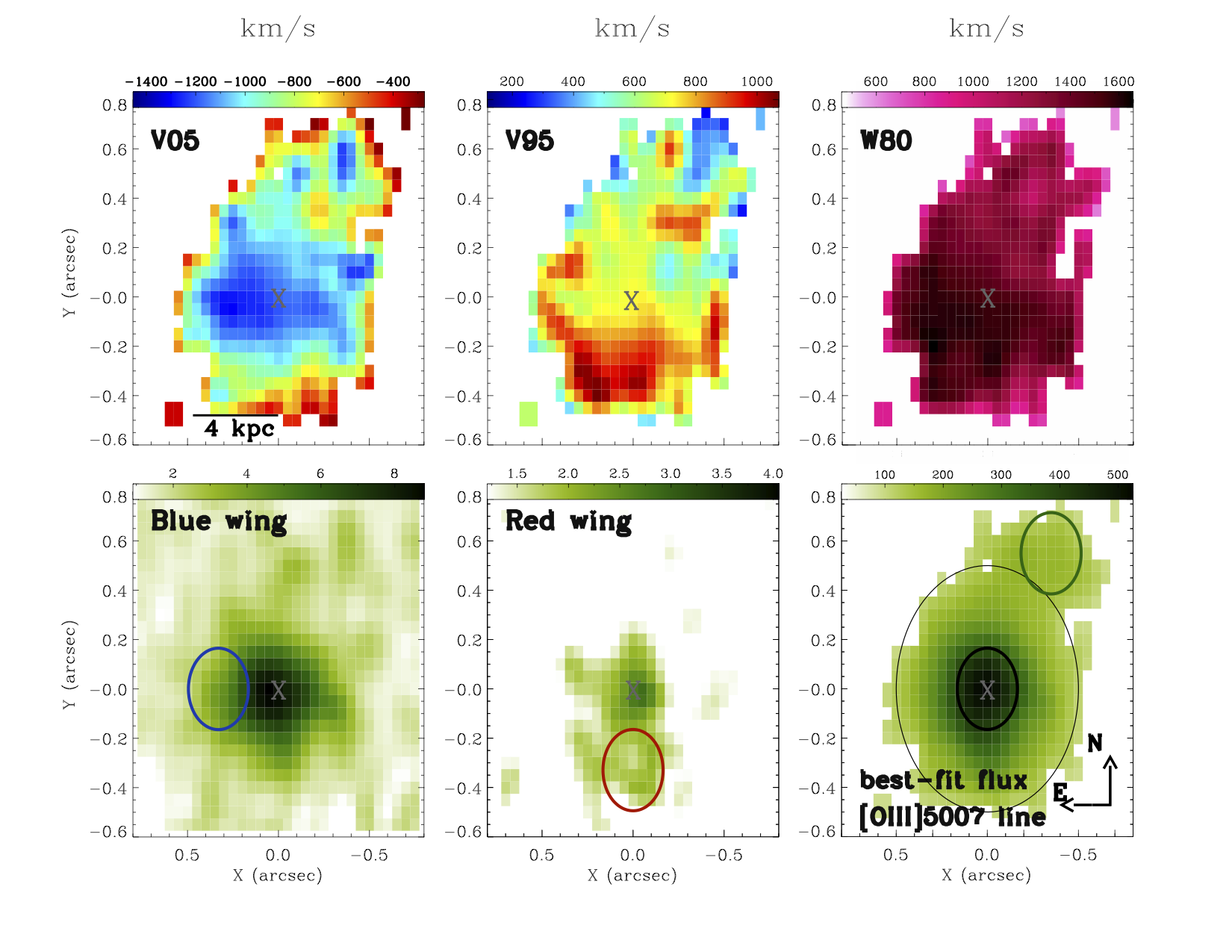}
  \includegraphics[width=14.5cm,height=9cm,angle=0]{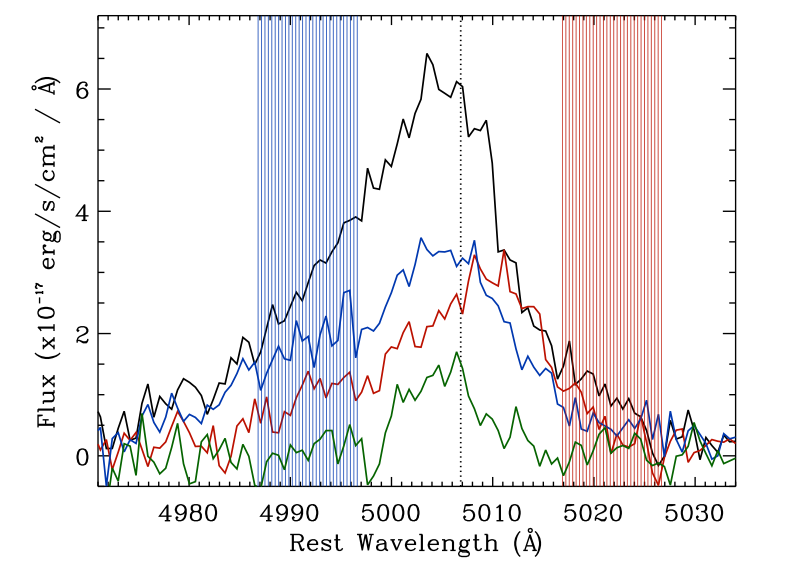}
      \caption{{\it Top panels}: Maps across the field of view of maximum blueshifted velocity (v$_{05}$), maximum redshifted velocity (v$_{95}$), and line width (w80). Blue-(red)~shifted velocities as high as V$_{\rm max, blue}\sim$-1400 km/s (V$_{\rm max, red}\sim$+1100 km/s) are found at distances as high as 4 kpc from the centre (indicated with a cross). 
{\it Central panels}: SINFONI [O III]5007 maps of the blue (-1200$\div$-600 km/s), red (600$\div$1200 km/s), and total [O III] flux (all scales are in units of 10$^{-15}$ erg cm$^{-2}$ s$^{-1}$). 
In the last panel we also plot as a reference the N-E direction, the J-band PSF, and the 0.4\arcsec\ and 1\arcsec\ diameter regions around the nucleus (sse Section 4.1 and 4.2). 
{\it Bottom panels:} SINFONI J-band spectra of the [O III]5007 line extracted from three regions of the size of the PSF at the position of the nucleus (black line; from the black circle in the total flux map), of the region with maximum V$_{\rm max,red}$ values (red line; from the red circle in the red wing panel), of the region of maximum V$_{\rm max,blue}$ values (blue line; from the blue circle in the blue wing panel), and at the position of the plume (green line; from the green circle in the total flux map). 
 The shaded regions show the wavelength ranges over which the blue and red wing flux maps have been collapsed.
    }
         \label{vmaps}
   \end{figure*}

\begin{table*}
\footnotesize
\begin{minipage}[!h]{1\linewidth}
\centering
\caption{Emission line properties in the integrated spectra}
\begin{tabular}{lccccccccc}
\hline
 & coordinates & H & K & $f_{H\alpha}$ & $f_{[O III]}$ &$f_{[O II]3727}$ &  z & $\lambda$ & FWHM\\
\hline
\multicolumn{10}{c}{Nucleus (integrated 1\arcsec)} \\
Narrow Component (NC) &  10:02:58.39 +02:10:13.94 & 19.70 & 19.90 &106$\pm$13  & 151$\pm$7  &  37$\pm$2    & 1.4718$\pm$0.0001    &   12376.1    &    550$\pm$20\\
Outflow Component (OC)$^a$ &  " & " & " & 147$\pm$31 &  467$\pm$6  &  51$\pm$2     &   1.4695$\pm$0.0001  &  12352.9 & 1735$\pm$35\\
\hline
\multicolumn{10}{c}{NW plume (integrated 0.4\arcsec)} \\
Single Component &  10:02:58.42 +02:10:14.4 & 23.90 & $25.30 $ & 33$\pm3$ &  53$\pm6$ &   .... & 1.4711$\pm$ 0.0001 & 12372.5  & 530$\pm15$ \\
\hline
\multicolumn{10}{c}{SE source} \\
Single Component &  10:02:58.46 +02:10:13.2 &  22.06 & 21.38 & $<20$ & $<10$ &  ... & ... & .... & 500 \\
\hline
\hline
\end{tabular}
\label{spectralfit}
\end{minipage}
Notes: All fluxes in $10^{-17}$erg/s/cm$^2$; all velocities in km/s.  
$^a$ The centroid of the OC profile is blueshifted of $\Delta v =$-272$\pm$22 km/s.

\end{table*}

\subsection{Spatially resolved analysis}

The left panel of Fig.~\ref{lines_syst} shows the SINFONI map extracted in the spectral range $\lambda$=1236.2-1238.9 nm, corresponding to the range where we observed the peak of the  emission of the [O~III]5007 line in the nucleus spectrum, which is highlighted as a yellow histogram in the right panel of Fig.~\ref{oxygen} , with a width of 600 km s$^{-1}$. Similarly, in the right panel of Fig. 7, we report the map of the peak of the H$\alpha$ emission, at $\lambda$=1620.6-1624.0 nm, integrated over a channel range assuming the same width of 600 km s$^{-1}$ as for the [O III] peak. In both panels, we overplot the contours of the Subaru/IRCS H-band (blue) and K-band (magenta) emission taken from Figure~\ref{cutouts}.  

The [O III] peak emission is overall symmetric, except for a plume extending beyond 0.5\arcsec\ ($\sim4.5$ kpc) in the North-west direction. This feature is also seen,  at lower significance  in the H$\alpha$ emission (right panel of Fig.~\ref{lines_syst}) and traces the NW component seen the Subaru/IRCS H-band data (in which the H$\alpha$ is redshifted). We extracted a spectrum from a region of the size of the PSF coincident with the NW plume, and we measured all the line fluxes, which are reported in Table 1.

From Fig.~\ref{lines_syst}, it is clear that no [O III] and H$\alpha$ emission at the redshift of the system is detected at the position of the SE source (denoted with the magenta and blue contours in the bottom left part of the images). We can place a 3$\sigma$ upper limit on the [O III] and H$\alpha$ fluxes of  $10^{-16}$ and 2$\times10^{-16}$ erg cm$^{-2}$ s$^{-1}$, assuming the same redshift of the source and a FWHM of the emission lines of 500 km s$^{-1}$. 
We searched for emission line features in the entire H- and J-band SINFONI datacubes and could not assign any redshift to the SE source.

\subsection{Kinematics analysis}\label{kin_an}

To map the line emission distributions and the corresponding velocities, we  analysed  the [O III] emission lines of each spaxel in the field of view separately. When the S/N is high enough, the complex [O III] profiles could be decomposed in multiple Gaussian components, while this is not the case where the S/N is lower. We defined the S/N as the ratio between the flux of the best-fit [O III]5007 total profile (or the flux of a Gaussian profile, characterised by a width equal to the FWHM of the NC of the 1$\arcsec$ spectrum, $\sim550km/s$) and  an amplitude equal to the standard deviation of the continuum regions in the vicinity of the emission line. In order to study variations of the line profiles across the field in the datacube, the non-parametric approach was used. This, in fact, does not depend on the number of Gaussian components used in each spaxel spectrum, thus also allowing a characterisation of the kinematics in spaxels at the lowest S/N ratio. However, we used an adaptive binning technique such that the high S/N regions were analysed over smaller areas than the low S/N regions. This technique  can be summarised as follows: as a first step, we chose to select only those spatial pixels (1x1 pixels) with a S/N ratio equal or higher than 2. When the ratio was lower than this threshold, we tried to increase it integrating the spectrum over larger areas (1x2 and, eventually, 2x2 pixel blocks). Finally, we chose all the pixel blocks with a S/N over the threshold value. 

 We measured, for the [O III] line, the same parameters extracted for the integrated [O II] and [O III] profiles (see section 2.1.1). 
The maps across the field of view of  the maximum blueshifted  velocity (V$_{\rm max, blue}$), the maximum redshifted velocity (V$_{\rm max, red}$), and w80 are shown in the first three panels of Fig.~\ref{vmaps}. Blue- (red)~shifted velocities with values as high as V$_{\rm max, blue}\sim$-1400 km/s (V$_{\rm max, red}\sim$+1100 km/s) are found, and the line width ranges from w80$\sim$800 to $\sim$1600 km s$^{-1}$.
The middle panels of Fig.~\ref{vmaps} show the SINFONI [O III]5007 flux maps of the blue (-1200$\div$-600 km/s; left), red (600$\div$1200 km/s; centre), and total (right) [O III] intensity. 
In the last panel of Fig.~\ref{vmaps} we show the [O III]5007 line from spectra extracted from three regions of the size of the PSF at the positions indicated in the central panels and colour coded accordingly: black from the nucleus (black circle in the total flux map), red from the region of maximum V95  values (red circle in the red wing panel), and blue from the region of maximum V05 values (blue circle in the blue wing panel). The shaded regions show the wavelength ranges over which the blue and red wing flux maps have been collapsed. 
The w80 measured for the spectra extracted at the maximum velocity regions, at distances $\sim2-3$ kpc from the centre, have large values ($\sim2000$ km s$^{-1}$), which cannot be ascribed to normal gravitational motions of gas in the disk of a galaxy. 

\begin{figure}
 \centering
  \includegraphics[width=9.cm,angle=0]{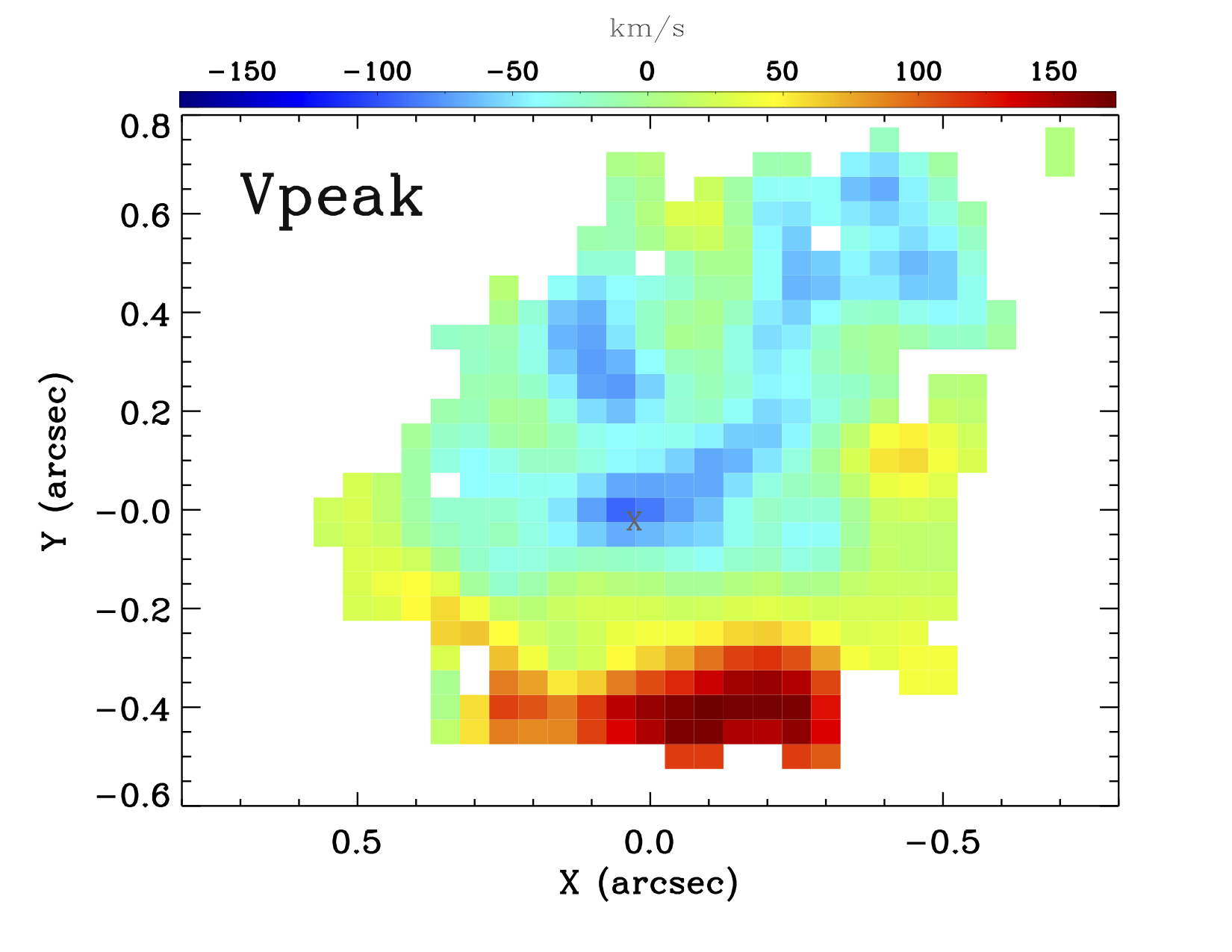} 
  \caption{Maps of the [O III] velocity peak (see Section 4.2) over the same FoV shown in Fig.~\ref{vmaps}. We measure a variation of Vpeak from -150 km s$^{-1}$ to +150 km s$^{-1}$, which we interpret as evidence for a rotation in the host galaxy. 
    }
         \label{vpeak}
   \end{figure}

Finally, Fig.~\ref{vpeak} shows the map of the [O III] velocity peak (Vpeak) obtained from our spaxels by spaxels fit with respect to the systemic emission defined in 4.1.
The results of a t-means test indicate that the velocities around the nucleus region and those in the reddest region have different means with a probability of false correlation P$<$0.01. Indeed, MonteCarlo analysis reveals that the non-parametric V$_{\rm peak}$ estimator is characterised by errors of the order of $\sim$ 50 km/s in the nuclear/top regions and of $\sim$ 80 km/s in the bottom regions. Taking  the errors into account,  the observed variation in V$_{\rm peak}$ across the FoV from -150 km/s to 150 km/s (a total of 300 km/s) is significant, and may represent the kinematic trace of a rotating disk in the host galaxy, although the gas kinematic appears perturbed by the outflow and/or by interactions.

%______________________________________________________________

%
%______________________________________________________________

%                                                One column figure
%----------------------------------------------------------- S_vib
   \begin{figure}
   \centering
  \includegraphics[width=9.cm,angle=0]{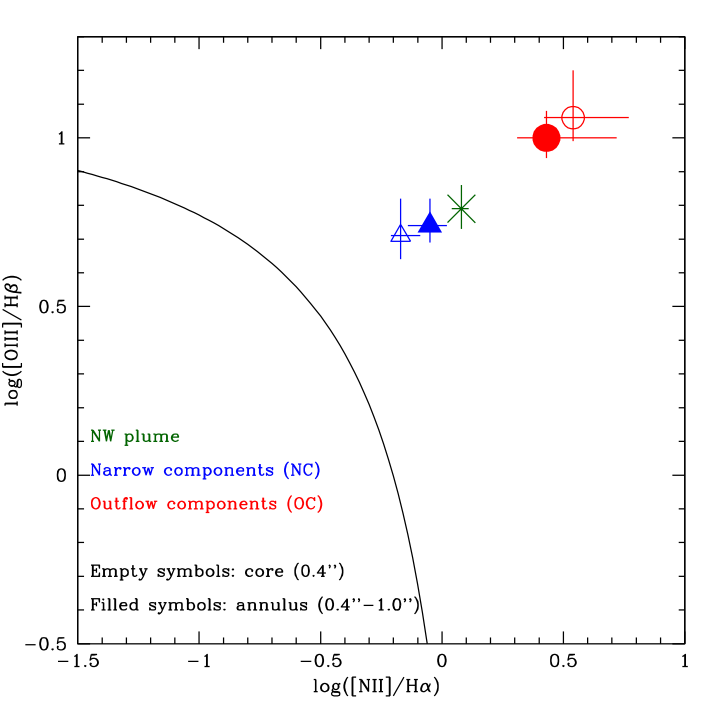} 
      \caption{Standard diagnostic diagram [O III]/H$\beta$-[NII]/H$\alpha$ with the theoretical z=1.47 curve
used to separate purely SF galaxies from galaxies containing
AGN (Kewley et al. 2013). 
The diagnostic ratios from the integrated 0.4\arcsec diameter nuclear spectrum (empty symbols) and for the 0.4-1\arcsec\ diameter annular region (filled symbols), for both the OC (red circles) and NC (blue triangles) components are shown. We also plot  the value obtained for the NW plume with a green cross.
    }
         \label{bpt}
   \end{figure}
%
%____________________________________________________

\subsection{Rest-frame optical flux ratios and ionisation source}\label{BPTsec}
We investigated the optical diagnostics routinely used to assess the photo-ionisation origin for the emission lines. In particular, we focused on the BPT diagram (from \citealt{Baldwin1981}) involving  the flux ratios [NII]/H$\alpha$ and [O III]/H$\beta$. The high resolution of the AO data, coupled with the achieved PSF, allowed us to study the ionisation mechanism in two separate regions: an inner region of the size of the PSF (0.4\arcsec/3.5 kpc diameter; inner black circle in the right panel of the central row in Fig.~\ref{vmaps}) corresponding to the inner $\sim9$ kpc$^{2}$, and an outer annular  region of 0.4 to 1\arcsec diameter (outer black circle in the right panel of the central row in Fig.~\ref{vmaps}), covering the 1.7 to 8.5 kpc radial distance from the centre, where the highest velocities are detected. 

We extracted two spectra separately from these two regions and we  fitted simultaneously, with the two kinematics component NC and OC described above, the [O III]+H${\beta}$ complex and the H${\alpha}$+[N II] complex to compute an excitation map of the galaxy. 
Figure~\ref{bpt} shows the BPT with the results from our  analysis. The solid line in the diagram corresponds to the theoretical redshift-dependent curve used to separate galaxies containing AGN from purely SF galaxies at z$\sim1.5$ \citep{Kewley2013}. In the figure we  plot  the diagnostic ratios obtained from the core 0.4\arcsec diameter spectrum (empty symbols) and from the 0.4\arcsec-1.0\arcsec\ diameter annular region (filled symbols), separately for the NC (blue triangles) and OC (red circles) components. In all cases the observed diagnostic ratios are consistent with an AGN photo-ionisation of the emission lines; all the data points lie above the \citet{Kewley2013} separation line, and similar conlusions are reached using the BPT diagnostics  involving  the [SII]/H$\alpha$ ratio (see also \citealt{Perna2015} for similar results in other two obscured QSOs at z$\sim1.5$). 

Mahony et al. (2015) report an increase of the ionisation state (higher [N II]/H$\alpha$ ratio) in the regions where the outflowing material is detected in 3C293, a local radio-loud QSO, which is interpreted as evidence of shocks associated with the wind (see also Section 5.4). Similarly, shock-ionisation models can predict values  of the [O III]/H$\beta$ ratio as high as $\sim10$ \citep{Allen2008}). We also report an increase of both line ratios for the OC component  with respect to the NC component, in both spectral extractions, which can be interpreted in the same way. 

Finally, In Fig.~\ref{bpt} we  plot  the flux ratios observed in the NW plume with a green cross. Also in this case, the ratios are consistent with a photo-ionisation origin from AGN emission. 

\section{Discussion}
We now discuss the interpretation of the velocity shifts revealed in the fits to the observed integrated [O III] line profiles, and of the high-resolution spatial and spectral analysis obtained through the adaptive optics data presented in the previous sections. We also relate the outflow properties with the X-ray, radio, and  host galaxy properties of XID5395.

\subsection{Outflow properties}
The most likely explanation for the OC component, revealed in the simultaneous fit of the VIMOS and SINFONI spectra as a broad and shifted extra-Gaussian component and resolved by the spatial analysis of the [O III] velocity maps, is the presence of an outflowing wind in this SB-QSO system.  
If this is the case, and assuming a biconical geometry, the kinetic power ($P_K$) and mass-outflow rate ($\dot M_{out}$) of the ionised component of the outflow can be computed under reasonable assumptions. In particular, estimates of the following quantities are needed: the electron temperature, $T_e$, and density, $n_e$, gas phase metallicity, emission line luminosity, spatial scale, and wind velocity. \\

\noindent 
{\it Electron Temperature, T$_{e}$}: The electron temperature can be calculated from the flux ratio  of two lines with the same  ionisation state.  For this purpose, we used the [O II]2471/[O II]3727 total flux ratio observed in the VIMOS spectrum and its  dependence on the electron temperature as found by  \citet{Humphrey2008} (their figure 7). 
We infer T$_e\approx13000\pm$500 K. We note that to correctly measure the electron temperature of the outflowing gas one should use the flux of the OC. Unfortunately, for both the faint [O II]2471 and blended [O II]3727 doublet it was not possible to constrain the OC and NC fluxes separately as for other emission lines (see section 4.1). Nevertheless, we postulate that the OC fluxes could contribute more than the NC fluxes (see Fig.~\ref{spectrum_sim}, OIII emission lines), making our estimate a good proxy of T$_e$.

\noindent  
{\it Electron Density, n$_{e}$}: An estimate of the electron density can be obtained from the flux ratio in the OC of the [SII]6716,6731 doublet, $R_{[SII]}=f([SII]6717)/f([SII]6731)$. Given that  the fitting procedure may produce strongly degenerate results when systemic components
are also present (see \citealt{Perna2015}), we used the value obtained from the fit to the [SII] doublet ratio in a 0.4-1.0\arcsec annulus centred on the target and covering only the south-east quadrant in which the outflow is observed. We obtained $R_{[SII]}=1\pm 0.1$.
The electron density can be estimated via the \citet{AckerJaschek95} formula, using the T$_{e}$ derived above. We obtain $n_e=780\pm300$cm$^{-3}$.

\noindent 
{\it Gas phase metallicity}: For our purposes, we assumed a solar metallicity, as the combination of emission lines available to us does not enable a clean calibration of the metallicity in the presence of strong AGN ionised gas. We note, however, that the assumed metallicity is consistent with that expected from the T$_{e}$ derived above, following the relation of \citet{Jones2015}. 

\noindent 
{\it Emission line luminosity}: Generally, the H$\beta$ luminosity is used to calculate the total amount of gas and mass outflow rate \citep[e.g.,][]{Liu2013,Cresci2015}. Alternatively, the H$\alpha$ flux can be adopted (see e.g. \citealt{Genzel2014}). In our case, the H$\beta$ line is faint, while the H$\alpha$ line may suffer from contamination from the hidden broad line region. We have therefore adopted the \citet{CanoDiaz2012} formalism, which instead employs the [O III] line luminosity.  We used the [O III]5007 flux associated with the outflow component in the 1\arcsec\ integrated spectrum (Figure \ref{spectrum_sim}, panel $a$).  As discussed in \citet{Carniani2015}, this approach gives most likely a lower limit to the total ionised component, given that in normal situations [O III] traces only a fraction of the gas with respect to H$\beta$. 

\noindent 
{\it Spatial scale}: We  adopt a radial extension of 4.3 kpc for the outflowing gas, given that we observe the bluewards emission out to this distance (see Figure \ref{vmaps}, panel V05, V95; Section 4.2).

\noindent 
{\it Outflow velocity}: Finally, we considered as outflow velocity the maximum velocity observed $v_{max}$ in the nuclear region ($v_{out}\approx 1300$ km/s; see Section 4.1), and we assumed that lower velocities are due to projection effects (\citealt{CanoDiaz2012, Cresci2015}).

With all these ingredients, following the Cano-D\'\i az et al.  (2012) formalism, we can derive the mass outflow rate and the outflow kinetic power, from the formulae below:

\begin{equation}\label{canodiaz}
P_K^{ion}=5.17\times 10^{43} \frac{CL_{44}([OIII])v_{out,3}^3}{n_{e3} R_{kpc} 10^{\left [ O/H \right ]}} erg\ s^{-1},
\end{equation}
\begin{equation}\label{Mcanodiaz}
\dot M_{out}^{ion}=164\times  \frac{CL_{44}([OIII])v_{out,3}}{n_{e3} R_{kpc} 10^{\left [ O/H \right ]}} M_\odot\ s^{-1},
\end{equation}

where $L_{44}([OIII])$ is the [OIII] luminosity associated with the outflow component in units of 10$^{44}$ erg s$^{-1}$, $n_{e3}$ is the electron density in units of 1000 cm$^{-3}$,  $v_{out,3}$ is the outflow velocity $v_{out}$ in unit of 1000 km s$^{-1}$,  $C$ is the condensation factor (assumed to be $\approx$ 1), 10$^{[O/H]}$ is the metallicity in solar units, and $R_{kpc}$ is the radius of the outflowing region in units of kpc. 

Using the values discussed above, we derive for XID5395 a kinetic power $P_k^{ion}\sim$2$\times$ 10$^{43}$ erg s$^{-1}$ and a mass outflow rate $\dot M\sim$45 $M_\odot/yr$. These values are  consistent overall with those observed for targets at similar bolometric luminosities \citep{Carniani2015,Stern2015_out}. These equations assume a simplified model where the wind occurs in a biconical region uniformly filled with outflowing ionised clouds. These values, not corrected for the extinction and regarding only the ionised component of the outflow (but see also the other conservative conditions in \citealt{CanoDiaz2012}, and the discussion in \citealt{Perna2015}, section 6.1), represent lower limits to the total outflow power. 

The inferred $\dot M$ ($>$45 $M_\odot/yr$) is lower than the  observed SFR ($\sim370$ $M_\odot/yr$), which can therefore in principle sustain the wind \citep{Veilleux2005}.
We note,  however, that the ionisation diagnostics exclude SF all over the FoV (See~\ref{BPTsec}) as an origin for the photo-ionisation of the gas. Considering also that the mass outflow rate refers only to the ionised component, while there are indications that the total molecular component may be up to a factor of 10 larger (see e.g. \citealt{Carniani2015}), and that we observe very high velocities ($>1300$ km s$^{-1}$), we can conclude that a SF-driven mechanism is unlikely to sustain entirely the outflow. 
Moreover, the derived kinetic power associated with the ionised gas is $\approx 0.4\%$ of the AGN bolometric luminosity  inferred from the SED fitting decomposition (logL$_{\rm bol}$=45.93), therefore, this power is consistent with the predictions of feedback models invoking AGN-driven winds (total kinetic power $\sim$ few \% of $L_{bol}$ ; \citealt{King2005}).

\subsection{Overall geometry and intrinsic properties}

The maps tracing the maximum blueshifted and redshifted velocities (V05 and V95 panels in Fig.~\ref{vmaps}) reveal that these high-velocity components are not spatially coincident with the core, i.e. they are displaced by $\sim$ 1 PSF size from the central AGN position, roughly in the east (blueshifted component) and south-east (redshifted component) directions. The outflow is seen mostly in this quadrant, and extends up to $\sim4.3$ kpc from the nucleus ($\sim0.5$\arcsec). 

When looking at the total fluxes associated with the fastest components, we see that the bulk of the approaching outflowing gas has an almost symmetric distribution overall the entire 1\arcsec\ diameter region ("blue wing" panel in Fig.~\ref{vmaps}); the receding gas, instead, is present only in the core and along the south direction ("red wing" panel in Fig.~\ref{vmaps}). These projected geometries are consistent with a wide-angle conical outflow, originating from the nucleus and moving towards our line of sight. 

\begin{figure}
   \centering
  \includegraphics[width=9.4cm]{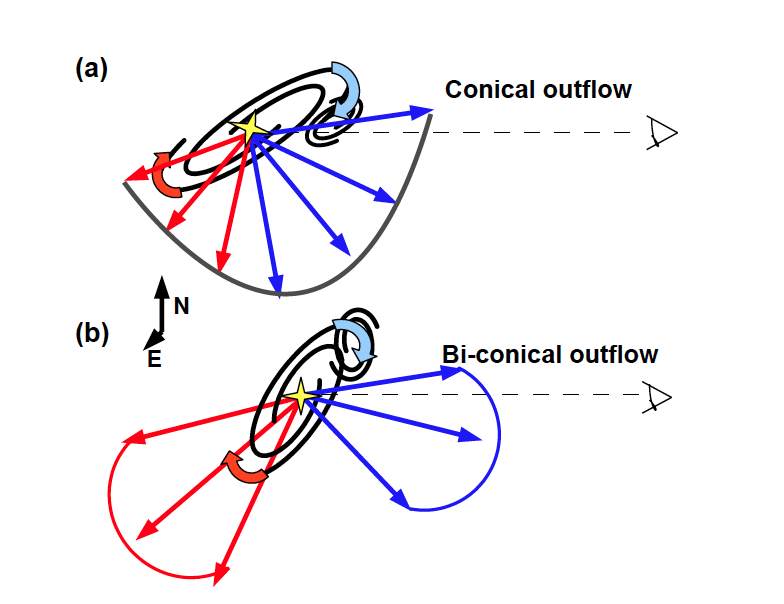}
   \caption{A cartoon illustrating the geometry of the system associated with XID5395, in the conical {\it (panel a)} and bi-conical {\it(panel b)} cases, as inferred from our spatial and spectral analysis. The line of sight intercepts the disk of the galaxy, which is rotating and inclined with respect the plane of the sky, so that the approaching part is in the North direction and the receding part is in the South direction, as revealed by the Vpeak map.  The (bi)-conical outflow is shown with arrows: blue arrows mark the gas moving towards our direction, and the  red arrows mark the gas moving opposite to our line of sight. See text for details. 
}
              \label{cartoon}%
    \end{figure}

Fig.~\ref{cartoon}a shows a schematic cartoon of the system,  illustrating the host, including the plume in the NW direction, and our inferred geometry for the wide-angle conical outflow. 
The outflowing region is located between the observer and the host; both approaching and receding gas regions are associated with this unique conical component. While the bulk of the gas has negative velocities (up to -1400 km/s), the far-side of the cone may be responsible for the receding emission with (absolute) lower velocities due to projection effects, as we indeed observe (+1000 km s$^{-1}$).  

A biconical outflow as depicted in  Fig.~\ref{cartoon}b  may be also consistent with our data. The flux observed in the nucleus in the red channel is a factor of $\sim3$ lower than that in the blue channel (see the black curve in the bottom panel of Fig.~\ref{vmaps}), which is roughly consistent with the extinction derived from the SED fitting and Balmer ratio decrement (A$_V$=1). In this case, the outflow should be tilted to observe the receding part located beyond the disk in the south direction, which should only be visible at larger distances (larger angles). In both cases, we estimate an opening angle of the cone at least of $\pi$/2. 

Unresolved compact jets may inject energy into the gas, heat it, and drive the outflows, as observed in radio-loud systems \citep{Begelman1989,Wagner2013}. Alternatively,  radiative-driven winds can induce shocks in the host and therefore accelerate relativistic particles (e.g. \citealt{Zubovas2012}), which then can emit in the radio band.
 The conical/wide angle geometry inferred for XID5395 is more consistent with a radiatively wind scenario of an expanding bubble (see e.g. \citealt{Zakamska2016_radio}) rather than with a jet-driven outflow scenario, for example as  observed in local radio sources with jets and ionised gas outflows (e.g. \citealt{Cresci2015_MUSE,Mahony2016}). Indeed, although XID5395 has a radio power typical of radio-loud objects, from the radio data in hand  we cannot confirm the presence of a jet (see Section 2.1), which should  be in any case confined within the central 1 kpc (the resolution of the VLBA data). The observed excess radio emission with respect to the SFR in XID5395, which is seen over the scale of the entire galaxy and is co-spatial with the outflow, can  therefore be in principle ascribed to radio emission produced in the wind during the shock; this is  suggested by \citet{Zakamska2016_radio} on the basis of a statistical analysis of the correlation between the radio luminosity and the narrow line kinematics in radio-quiet AGN.
At luminosities higher than  L$_{\rm bol}>3\times10^{45}$ erg s$^{-1}$ (as the case of XID5395), the wind can in principle sustain and also generate high radio luminosities.

%                                     Two column figure (place early!)
%______________________________________________ Gamma_1 (lg rho, lg e)

\subsection{Outflows, starbursts, and luminous AGN}

\begin{figure}
   \centering
  \includegraphics[width=8.3cm]{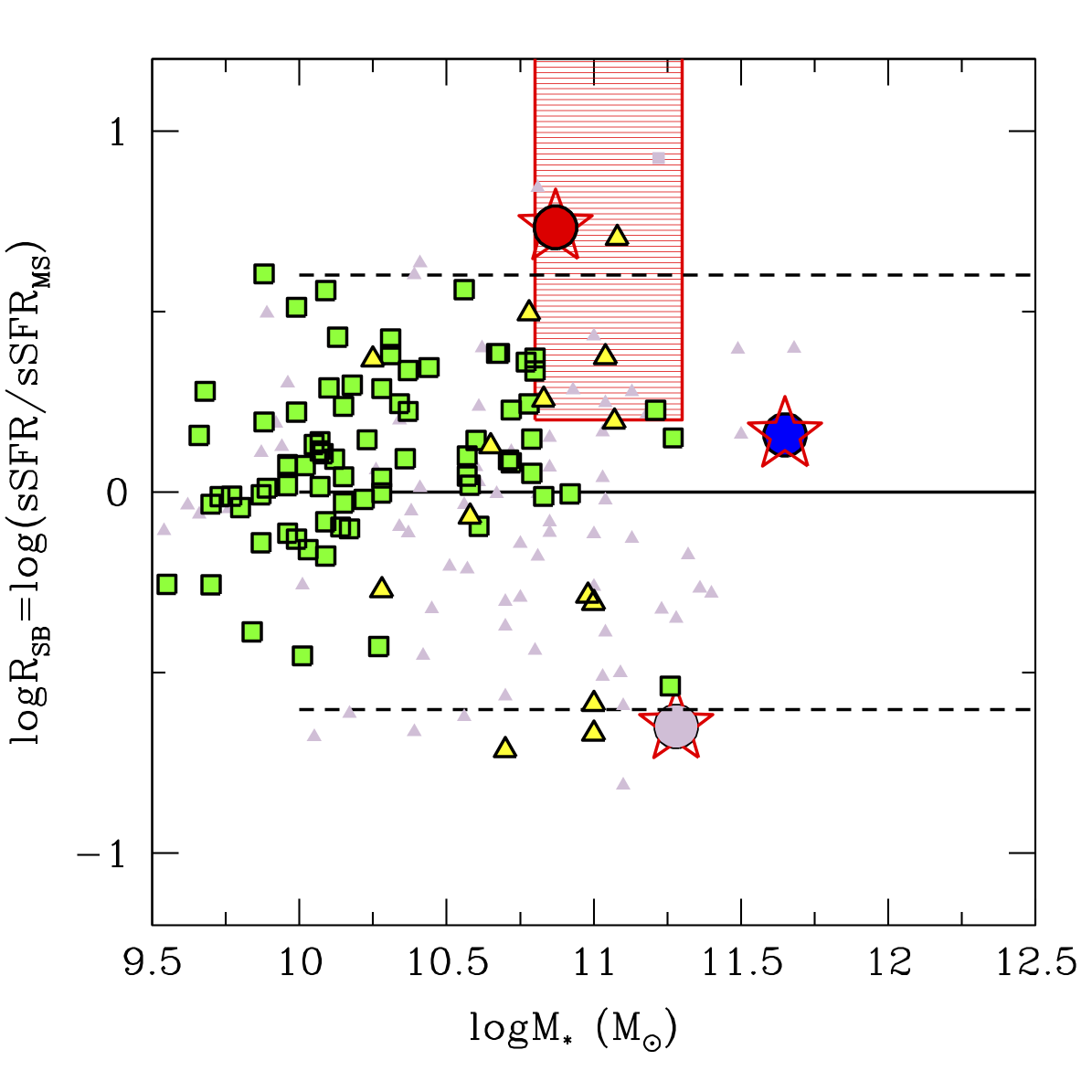}
      \caption{Logarithm of the starburstiness, R$_{\rm SB}$, defined as the ratio of the sSFR to that expected for MS galaxies as parametrised by Whitaker et al. (2012), as a function of the logarithm of the stellar mass. 
XID5395 is the red star. We also plot, with starred symbols, two additional sources from XMM-COSMOS for which the outflows have been revealed: XID2028 at z$\sim1.6$, from \citealt{Cresci2015}, blue; and XID70135/MIRO20581 at z$\sim2.5$ from \citealt{Perna2015_miro}, grey. Green squares are objects from the MASSIV survey \citep{Contini2012} in the redshift range z=1-2; yellow triangles are sources at z$=1-2$ in the \citet{Genzel2014} IFU sample. We also plot, with small grey triangles, all the other sources in the  \citet{Genzel2014} samples in the redshift range z=$2-3$.  The shaded red region indicates the range of sSFR and stellar mass of the targets in \citet{Harrison2012}. 
    }
         \label{rsb}
   \end{figure}

Figure~\ref{rsb} shows the position of XID5395 in the ``starburstiness" (R$_{\rm SB}$) versus stellar mass plane, where R$_{\rm SB}$ is defined as the ratio of the specific SFR (sSFR=SFR/M$_\star$) to that expected for MS galaxies (parametrised by \citealt{Whitaker2012}). XID5395 is shown as a red starred circle: from the host galaxy properties the source is indeed classified as ``starburst" (e.g. a factor of $>4$ above the MS, following the definition of \citealt{Rodighiero2011}). In Figure~\ref{rsb} we also plot, for comparison, a compilation of objects at z$=1-2$ for which SINFONI data are available, $\sim90$  targets  observed within various programmes \citep{Mancini2011,Contini2012,Genzel2014}: SINS (Spectroscopic Imaging survey in the Near-infrared with SINFONI; yellow triangles), SINS/zC-SINF (the continuation of SINS in the COSMOS field; yellow triangles), and MASSIV (Mass Assembly Survey with SINFONI in VVDS; green squares). 

In the ``starburstiness" vs. stellar mass plot, XID5395 lies in the same locus occupied by a sample of eight QSOs at z$\sim2.5$, originally selected as submillimetre galaxies, with kpc-scales outflow detected by NIR IFU data (\citealt{Harrison2012}; red shaded area). On the contrary, the only source from the galaxy sample with the same extreme host properties of XID5395 (D3a-6397 reported in \citealt{Genzel2014}; yellow triangle above the ``starbursts" threshold in Figure~\ref{rsb}), shows a massive, rotating disk with no sign of outflow (see \citealt{Cresci2009}). Also, no AGN activity has been detected in this source, at least according to the BPT classification. 
\citet{Brusa2015} and \citet{Perna2015_miro}, on the basis of X-shooter and SINFONI follow-up observations of COSMOS red quasars at z$\sim2$, revealed the presence of  galaxy-scale ionised outflow (extending out to 4-10 kpc)  in non-starburst systems also.   
In Fig.~\ref{rsb} we plot the two luminous QSOs from the XMM-COSMOS sample with  detected outflow and with high-resolution SINFONI data:
XID2028 \citep[starred blue circle]{Cresci2015} and MIRO20581 \citep[starred grey circle]{Perna2015_miro}. When considering all the luminous XMM-COSMOS sources with detected outflows altogether (XID5395, XID2028, and MIRO20581), all of them are in the massive tail (M$_*=8\times10^{10}-5\times10^{11}$ M$_\odot$) of the MS galaxy, but  have clearly different star-forming properties, spanning the entire range from below to above the MS. 

The fact that the outflow presence does not depend on the host galaxy properties (in particular the starburstiness), but is rather associated with the presence of AGN in massive systems corroborates the results reported in \citet{FS2014} and \citet{Genzel2014}, based on the detailed analysis of H$\alpha$ emission in NIR stacked spectra of star-forming galaxies at z$\sim2$ (see also similar results for less luminous systems presented in \citealt{Cimatti2013}). Moreover, comparing XID5395 and D3a-6397 discussed above (e.g. both are starburst galaxies), we see a clear outflow in the galaxy with a strong AGN. This may be seen as further indication that the ``starburstiness'' can be a good method to select objects in the outflowing phase only  when coupled with the presence of a luminous AGN. 
The luminous AGNs in which we detect the outflow in the individual spectra (starred symbols in Figure~\ref{rsb}), with L$_{\rm bol}>10^{46}$ erg s$^{-1}$, are the tip of the iceberg of the overall population hosting AGN, and because they are more luminous they obviously show the features with higher statistics. Moreover, given that we were able to spatially map the [O III] emission instead of the H$\alpha$ region, our results are less affected by possible contamination from the BLR emission (see also discussion in \citealt{Perna2015_miro}).

The outflow presence also does not seem to correlate with the nuclear obscuration revealed in the X-rays. The three XMM-COSMOS sources with detected outflows, selected in different ways, show column densities ranging from N$_{\rm H}\sim5\times10^{21}$ cm$^{-2}$ (mild obscuration) to N$_{\rm H}\sim5\times10^{23}$ cm$^{-2}$ (heavy obscuration; see also \citealt{Harrison2016} for a lack of correlation of the outflows on the nuclear obscuration).  The only common feature is a considerable extinction in the optical band (e.g. they are all classified as red/reddened AGN) and an intrinsic X-ray luminosity that is higher than $>10^{44.5}$ erg s$^{-1}$. Although the high X--ray luminosity can be ascribed to a selection effect, all these sources show a contribution of the X--ray emission to the bolometric luminosity that is on average higher than that observed for classic blue quasars of similar bolometric luminosity; we measure  k$_{\rm bol}\sim$10 for outflowing sources, while  \citet{Lusso2012} reported k$_{\rm bol}\sim$25 for typical X-ray selected Type 1 QSOs of L$_{\rm bol}\sim10^{46}$ erg s$^{-1}$. 
Such low bolometric corrections (higher contribution of L$_X$ to L$_{\rm bol}$) may be related to a balance of disk (directly related to L$_{\rm bol}$) and corona (directly related to L$_{\rm X}$) emissions that are different from what is observed in typical blue, unobscured QSO. This can in turn be ascribed to the very processes governing the accretion and ejection mechanisms of material in the accretion disk region during the short-lived wind phase. 

%                                                One column figure
%----------------------------------------------------------- S_vib

\subsection{On the transition nature of XID5395} 

From our multi-wavelength and spectroscopic analysis, we have a clear picture of the XID5395 system, which can be summarised as follows. We have a luminous AGN (L$_{\rm bol}\sim8\times 10^{45}$ erg s$^{-1}$), obscured in the X-rays (N$_{\rm H}\sim10^{23}$ cm$^{-2}$) and at optical wavelengths (A$_V\sim1$ from Balmer decrement and SED fitting; see Section 4.1), hosted in a massive (M$_*\sim7.8\times10^{10}$ M$_\odot$), starburst (SFR$\sim$370 M$\odot$ yr$^{-1}$) disk galaxy, possibly interacting with a lower luminosity system. A wide-angle outflow of ionised gas is launched  from the central core at high velocities ($>1300$ km s$^{-1}$), and it is propagating in the ISM in the direction of our line of sight.

The fact that vigorous SF  and AGN activity do co-exist in XID5395 can be seen as a first indication of the transition nature of this source. Indeed, in the co-evolutionary models of galaxy-AGN evolution (e.g. \citealt{Hopkins2008}) the phase of maximum and co-eval SF and accretion rate last only $\sim$few$\times10^{8}$ yr, and is possibly triggered by mergers. The presence of the interacting galaxy seems to confirm that we caught XID5395 exactly during this crucial and short phase. 

Given the steep spectral index observed at 1.4 and 3 GHz ($\alpha\sim$1.1 - 1.3; \citealt{Schinnerer2010}; Smol\v{c}i\'c et al., in preparation), XID5395 can be classified as a compact steep spectrum radio source (CSS; see \citealt{Orienti2015}) with a size comparable to that of the host galaxy ($\sim$few kpc).  Compact radio sources at high-z  may be intrinsically young sources (e.g. \citealt{Fanti1995}) in which a dense ISM still prevents the radio emission from expanding at larger scales, which   likely evolve over longer timescales into the giant FR I radio galaxies observed at low redshifts. This is a further indication that XID5395 may be caught in an early stage of its evolution.

Young radio sources and CSS also show disturbed kinematics in the ionised gas (e.g. \citealt{Gelderman1994}) with exceptionally broad widths observed in the [O III] lines \citep{Kim2013}.  An example of a local CSS source is 3C293 at z=0.045, where the inner jet is clearly resolved and confined within few kpc, while high-velocity  ionised gas is seen up to $\sim10$ kpc in IFU data \citep[see their Figure 7]{Mahony2016}. Moreover, the molecular and ionised gas components reveal different morphologies, which can be explained with the presence of a spherical-like bubble as predicted in simulations (e.g.\citealt{Wagner2013}; see also discussion in Section 5.2) inflated by the jet itself, but independently expanding in the ISM.
In addition to the exceptional broad line widths, 3C293 has other common properties with XID5395: both sources show similar stellar masses ($\sim10^{11}$ M$\odot$), red colours and disturbed morphology, indicating the presence of a companion galaxy  that can be the remnant of a merger \citep{Floyd2008,Lanz2015}. 
XID5395 can therefore be considered the high-redshift analogue of 3C293, in which the same phenomenon of inflating a spherical-like bubble of gas by an inner jet may be in place. However, we cannot resolve any jet due to resolution limits of the observations, although we have evidence of the presence of  a bubble of ionised gas and diffuse radio emission, with a wide opening angle. In fact, as proposed in Section 5.2, the radio diffuse emission can be ascribed to shocks in the dense ISM \citep{Zakamska2016_radio}, suggesting that XID5395 is in a phase of radiative feedback. 

\begin{figure}
   \centering
  \includegraphics[width=9.4cm]{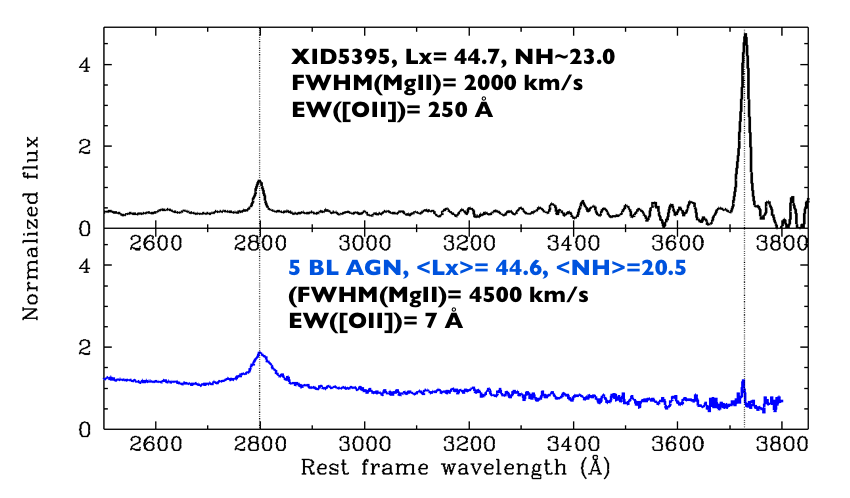}
   \caption{Rest-frame  VIMOS spectrum of  XID5395 (top) and the average VIMOS spectrum of five unobscured BL AGN (bottom) chosen in the same z and L  bins of our target. In both panels, the dotted lines indicate the position of the MgII2800 (left) and [O II]3727 (right) lines.  The y-axis scale is maintained in the two panels  to ease a direct comparison of the two spectra; it is clear that the [O II] emission weakens, while the continuum becomes bluer and the MgII line becomes stronger in the bottom panel. 
}
              \label{spectrum}%
    \end{figure}

If the source is caught in this short transition phase, in few hundreds Myrs the SF will be likely completely suppressed or inhibited, and the source will evolve into an unobscured QSO. 
In Figure~\ref{spectrum}, we plot the rest-frame zCOSMOS VIMOS spectrum over the full wavelength range of XID5395 (upper panel) compared to  the average spectrum of five unobscured BL AGN at the same average  $<z>$ and of the same intrinsic $<$L$_{\rm X}>$ (L$_{\rm X}\sim7\times10^{44}$ erg s$^{-1}$, bottom panel). From the comparison of the two spectra, we see the emergence of the broad MgII line (from FWHM$\sim2000$ km s$^{-1}$ as observed in XID5395, to FWHM$\sim4000$ km s$^{-1}$, as measured in the average BL AGN spectrum), with the continuum moving from flat to blue, and the [O II] line; this line dominates in the spectrum of XID5395, weakening in the BLAGN spectrum, as expected if SF is completely suppressed at the end of the co-evolutionary sequence. Red  sources with flat rest-frame UV continuum, as observed in XID5395, have been historically missed in optical surveys of QSO, but gained increasing attention in recent years as signposts of objects in the blow-out phase (see e.g. \citealt{Urrutia2008,Brusa2010,Banerji2012,Brusa2015,Zakamska2016_xshooter} among others).

Finally, in XID5395 we measure a momentum flux,  $\dot P_k^{\ ion}$=$\dot M v_{out}\sim$3.5$\times$10$^{35}$ dyne, $\approx$$L_{bol}/c$. Although this quantity should be considered a lower limit because it only traces the ionised gas component, it is among the lowest values reported so far for similar systems studied in the [O III] outflows. \citet{Stern2015_out} derived a theoretical upper limit on the instantaneous force due to hot gas pressure acting on the outflows of 7$\times$ $L_{bol}/c$, where $L_{bol}/c$ traces the radiative momentum flux from the central black hole. Previous studies on the momentum flux of both low-z and high-z AGN (see e.g. the compilation in \citealt{Carniani2015}) report larger values for the time-averaged estimate of this quantity ($\sim20$).  As suggested by \citet{Stern2015_out}, the discrepancy between the measured momentum flux (e.g. time-averaged force) and the predicted instantaneous force in the samples studied so far can be explained with a force decreasing with time; this is consistent with models in which initially the accretion is  fully obscured and the hot gas resulting from the accretion disk wind is confined by the ambient ISM, until feedback clears the nucleus and the radiative pressure drops.  The fact that we observe a low value of value of $\dot P_k^{\ ion}$ may further support the young/transition nature of this source. 

%                                     Two column figure (place early!)
%______________________________________________ Gamma_1 (lg rho, lg e)

\section{Summary}
We exploited VLT/VIMOS, VLT/SINFONI, and Subaru/IRCS AO data to study the kinematics properties of an X-ray luminous, obscured  SB-QSO system at z$\sim1.5$, associated with an extreme EW of the [O II] line (EW$\sim200$ \AA; see Fig.~\ref{selection}). 

We propose that EW larger than 150 \AA,\, when coupled with an X-ray emitting source, may be a very efficient criterion to isolate objects with disturbed morphology  (Fig.~\ref{ACS}),  and possibly associated with an outflowing phase, as witnessed by the broad, asymmetric line profiles of the [O II] and [O III] emission lines in the integrated VIMOS and SINFONI spectra of the target presented in this study (Fig.~\ref{oxygen}).
This criterion may complement other diagnostics that have been proposed in the past to isolate such rare objects in X-ray samples, such as those based on the observed red colours of the hosts (e.g. \citealt{Brusa2015,Perna2015_miro}).  

The kinematics analysis, coupled with the high spatial resolution achieved through AO observations (few kpc), reveals complexities and asymmetries in and around the nucleus of XID5395, which are summarised below: 
   \begin{enumerate} 
      \item SUBARU/IRCS data reveal three distinct sources in the proximity of XID5395 (Fig.~\ref{cutouts}): a central source with a projected transversal size of $\sim$8.5 kpc (``nucleus") that is associated with the X-ray source; a compact object at about 10 kpc towards the SE direction (``SE source"); an extended source at $\sim$4 kpc towards the NW direction (``NW plume");
 \item Within the nucleus, the velocity field measured in the SINFONI data via multicomponent spectral fitting and non-parametric analysis reveals different kinematic components (see the NC and OC components in Fig.~\ref{spectrum_sim} and the velocity maps in Fig.~\ref{vmaps}); in particular, the maximum blueshifted and redshifted velocities, with absolute values $\gsimeq900-1300$ km s$^{-1}$,  are not spatially coincident with the nuclear core (the central PSF, 0.4\arcsec). 
In addition, from the variation of the velocity peaks within the nucleus, we  detect signatures of a rotating disk (Fig.~\ref{vpeak});
\item SINFONI data confirm that the NW plume traces an object, possibly a merging galaxy, at the same redshift of the central AGN (see Fig.~\ref{lines_syst}) with lower velocity components (e.g. w80$<1000$ km s$^{-1}$; see Fig.~\ref{vmaps}) than those reported for the nucleus. Instead, we cannot confirm,  nor exclude, that the SE component is at the same redshift as the nucleus, and part of an interacting system at a pre-merger phase, or right after a first pass in the merger path. If this were the case, the SE component should be at least a factor of 10 less luminous/massive;
 \item The BPT diagram suggests that AGN photo-ionisation is responsible for the line emission ratios observed both in the NW plume and over the entire nucleus (Fig.~\ref{bpt}) with a possible contribution of shocks towards the external regions.  
\item We best model the observed emission as depicted in Fig.~\ref{cartoon}: Our line of sight intercepts a wide-angle conical and fast outflow, which originates from the central AGN (Sect.~5.2). The galaxy disk is tilted with respect the outflow cone to observe both the blue and redshifted emissions. The kinetic power, mass outflow rate, and momentum rate inferred from our analysis are in the range expected from AGN feedback models originated by radiatively driven winds (Sect.~5.1);  
\item Although this proposal is still on the basis of a small sample, we propose that outflowing QSO may be associated with objects with a higher contribution of L$_X$ to L$_{\rm bol}$ than the average observed in typical blue, unobscured QSOs (Sect.~5.3); 
\item Overall, we accumulated different observational evidence (presence of outflow and merger, co-eval SF and AGN activity, and compact radio emission; all described in Section~5.4) that XID5395 is caught in the short, transition phase of ``feedback" in the AGN-galaxy co-evolutionary path, and it will subsequently evolve into an unobscured QSO (see Fig.~\ref{spectrum}). 
   \end{enumerate}

From the analyses of both galaxies (e.g. \citealt{Cimatti2013,Genzel2014}) and AGN (e.g. \citealt{Harrison2016}) samples, it is becoming clear that powerful outflows may be very common in AGN hosts, especially at high redshifts (see e.g. \citealt{Marziani2016}).   However,  to uncover the kinematics and physical properties of the wind,  high-resolution and high S/N observations, such as those presented in this paper, are needed.  In the next few years the ESO Large Programme SUPER (a {\tt SINFONI} Survey for Unveiling the Physics and Evolution of Radiative feedback, PI: V. Mainieri) will undertake this task by targeting, with the needed resolution and depth, a sample of $\sim$40 AGN at z$\sim$2.
In addition, multiple diagnostics should be employed to select the best cases and to study the effect of feedback from AGN in action in high-z sources.  Indeed, the correspondence between luminous red sources and "feedback phase with prominent outflows" may not be exclusive, because the physical conditions  that determine the presence and observability of an outflow (e.g. the wind velocity and momentum rate, fraction of energy released, properties in the ISM in which the outflows propagate) {\it and} the observed host galaxies properties of AGN (e.g. the timescales of accretion and star formation processes; the viewing angle) may vary from source to source. This translates into the fact that galaxy-wide outflows can also be observed in blue and unobscured QSOs (e.g. \citealt{Harrison2016}), especially at the highest luminosity (e.g. \citealt{Carniani2015}), for which the line of sight may be cleaner for the combination of both viewing angle and larger energetics. Significant advances on this topic will come with the analysis of large and unbiased samples of infrared spectra of AGN selected from multi-wavelength surveys (e.g. SUPER, KASHz surveys). 
Finally, The molecular component of the outflows is still to be explored in detail, and follow-up observations with NOEMA and ALMA are needed to obtain the full picture on the energetics and processes that dominate the outflows.

\begin{acknowledgements}
Based on observations undertaken at the European Southern Observatory, Paranal, Chile under the programme 094.A-0250 and the Large Programme 175.A-0839.  
Based on data collected at Subaru Telescope, which is operated by the National Astronomical Observatory of Japan.
MP, MB, and GL acknowledge support from the FP7 Career Integration Grant ``eEASy'' (``SMBH evolution through cosmic time: from current surveys to eROSITA-Euclid AGN Synergies", CIG 321913).  ID acknowledges the European Union's Seventh Framework programme under grant  agreement 337595 (ERC Starting Grant, ``CoSMass").
%MB gratefully acknowledges fundings from the DFG cluster of excellence `Origin and Structure of the Universe' (www.universe-cluster.de).
We acknowledge financial support from INAF under the contracts PRIN-INAF-2011 (``Black Hole growth and AGN feedback through cosmic time"), PRIN MIUR 2010-2011 (``The dark Universe and the cosmic evolution of baryons"), and PRIN INAF 2014 (``Windy Black Holes combing galaxy evolution"). 
We gratefully acknowledge the unique contribution of the entire COSMOS collaboration for making their excellent data products publicly available; more information on the COSMOS survey is available at \verb+http://www.astro.caltech.edu/~cosmos+. In particular, we thank Micol Bolzonella, Paolo Ciliegi, Francesca Civano, Stefano Marchesi, Vernesa Smol\v{c}i\'c, and Noelia Herrera Ruiz for useful discussions and for sharing data prior to publication. We thank the anonymous referee for his/her constructive comments to the paper. 
\end{acknowledgements}

% WARNING
%-------------------------------------------------------------------
% Please note that we have included the references to the file aa.dem in
% order to compile it, but we ask you to:
%
% - use BibTeX with the regular commands:
%   \bibliographystyle{aa} % style aa.bst
%   \bibliography{Yourfile} % your references Yourfile.bib
%
% - join the .bib files when you upload your source files
%-------------------------------------------------------------------

%\bibliographystyle{aa} % style aa.bst
 %\bibliography{angi.bib} % your references Yourfile.bib

\begin{thebibliography}{106}
\expandafter\ifx\csname natexlab\endcsname\relax\def\natexlab#1{#1}\fi

\bibitem[{{Acker} \& {Jaschek}(1995)}]{AckerJaschek95}Acker, A., \& Jaschek, C. 1995, Sci, 270, 1236J

\bibitem[{{Alexander} {et~al.}(2010){Alexander}, {Swinbank}, {Smail},
  {McDermid}, \& {Nesvadba}}]{Alexander2010}
{Alexander}, D.~M., {Swinbank}, A.~M., {Smail}, I., {McDermid}, R., \&
  {Nesvadba}, N.~P.~H. 2010, \mnras, 402, 2211

\bibitem[{{Allen} {et~al.}(2008){Allen}, {Groves}, {Dopita}, {Sutherland}, \&
  {Kewley}}]{Allen2008}
{Allen}, M.~G., {Groves}, B.~A., {Dopita}, M.~A., {Sutherland}, R.~S., \&
  {Kewley}, L.~J. 2008, \apjs, 178, 20

\bibitem[{{Amor{\'{\i}}n} {et~al.}(2015){Amor{\'{\i}}n}, {P{\'e}rez-Montero},
  {Contini}, {V{\'{\i}}lchez}, {Bolzonella}, {Tasca}, {Lamareille}, {Zamorani},
  {Maier}, {Carollo}, {Kneib}, {Le F{\`e}vre}, {Lilly}, {Mainieri}, {Renzini},
  {Scodeggio}, {Bardelli}, {Bongiorno}, {Caputi}, {Cucciati}, {de la Torre},
  {de Ravel}, {Franzetti}, {Garilli}, {Iovino}, {Kampczyk}, {Knobel}, {Kova{\v
  c}}, {Le Borgne}, {Le Brun}, {Mignoli}, {Pell{\`o}}, {Peng}, {Presotto},
  {Ricciardelli}, {Silverman}, {Tanaka}, {Tresse}, {Vergani}, \&
  {Zucca}}]{Amorin2015}
{Amor{\'{\i}}n}, R., {P{\'e}rez-Montero}, E., {Contini}, T., {et~al.} 2015,
  \aap, 578, A105

\bibitem[{{Aoki} {et~al.}(2005){Aoki}, {Kawaguchi}, \& {Ohta}}]{Aoki2005}
{Aoki}, K., {Kawaguchi}, T., \& {Ohta}, K. 2005, \apj, 618, 601

\bibitem[{{Aversa} {et~al.}(2015){Aversa}, {Lapi}, {de Zotti}, {Shankar}, \&
  {Danese}}]{Aversa2015}
{Aversa}, R., {Lapi}, A., {de Zotti}, G., {Shankar}, F., \& {Danese}, L. 2015,
  \apj, 810, 74

\bibitem[{{Bae} \& {Woo}(2014)}]{Bae2014}
{Bae}, H.-J. \& {Woo}, J.-H. 2014, \apj, 795, 30

\bibitem[{{Baldwin} {et~al.}(1981){Baldwin}, {Phillips}, \&
  {Terlevich}}]{Baldwin1981}
{Baldwin}, J.~A., {Phillips}, M.~M., \& {Terlevich}, R. 1981, \pasp, 93, 5

\bibitem[{{Banerji} {et~al.}(2012){Banerji}, {McMahon}, {Hewett},
  {Alaghband-Zadeh}, {Gonzalez-Solares}, {Venemans}, \&
  {Hawthorn}}]{Banerji2012}
{Banerji}, M., {McMahon}, R.~G., {Hewett}, P.~C., {et~al.} 2012, \mnras, 427,
  2275

\bibitem[{{Begelman} \& {Cioffi}(1989)}]{Begelman1989}
{Begelman}, M.~C. \& {Cioffi}, D.~F. 1989, \apjl, 345, L21

\bibitem[{{Bell}(2003)}]{Bell2003}
{Bell}, E.~F. 2003, \apj, 586, 794

\bibitem[{{Berta} {et~al.}(2013){Berta}, {Lutz}, {Santini}, {Wuyts}, {Rosario},
  {Brisbin}, {Cooray}, {Franceschini}, {Gruppioni}, {Hatziminaoglou}, {Hwang},
  {Le Floc'h}, {Magnelli}, {Nordon}, {Oliver}, {Page}, {Popesso}, {Pozzetti},
  {Pozzi}, {Riguccini}, {Rodighiero}, {Roseboom}, {Scott}, {Symeonidis},
  {Valtchanov}, {Viero}, \& {Wang}}]{Berta2013}
{Berta}, S., {Lutz}, D., {Santini}, P., {et~al.} 2013, \aap, 551, A100

\bibitem[{{Bongiorno} {et~al.}(2012){Bongiorno}, {Merloni}, {Brusa},
  {Magnelli}, {Salvato}, {Mignoli}, {Zamorani}, {Fiore}, {Rosario}, {Mainieri},
  {Hao}, {Comastri}, {Vignali}, {Balestra}, {Bardelli}, {Berta}, {Civano},
  {Kampczyk}, {Le Floc'h}, {Lusso}, {Lutz}, {Pozzetti}, {Pozzi}, {Riguccini},
  {Shankar}, \& {Silverman}}]{Bongiorno2012}
{Bongiorno}, A., {Merloni}, A., {Brusa}, M., {et~al.} 2012, \mnras, 427, 3103

\bibitem[{{Brandt} \& {Alexander}(2015)}]{Brandt2015}
{Brandt}, W.~N. \& {Alexander}, D.~M. 2015, \aapr, 23, 1

\bibitem[{{Brusa} {et~al.}(2015){Brusa}, {Bongiorno}, {Cresci}, {Perna},
  {Marconi}, {Mainieri}, {Maiolino}, {Salvato}, {Lusso}, {Santini}, {Comastri},
  {Fiore}, {Gilli}, {Franca}, {Lanzuisi}, {Lutz}, {Merloni}, {Mignoli},
  {Onori}, {Piconcelli}, {Rosario}, {Vignali}, \& {Zamorani}}]{Brusa2015}
{Brusa}, M., {Bongiorno}, A., {Cresci}, G., {et~al.} 2015, \mnras, 446, 2394

\bibitem[{{Brusa} {et~al.}(2010){Brusa}, {Civano}, {Comastri}, {Miyaji},
  {Salvato}, {Zamorani}, {Cappelluti}, {Fiore}, {Hasinger}, {Mainieri},
  {Merloni}, {Bongiorno}, {Capak}, {Elvis}, {Gilli}, {Hao}, {Jahnke},
  {Koekemoer}, {Ilbert}, {Le Floc'h}, {Lusso}, {Mignoli}, {Schinnerer},
  {Silverman}, {Treister}, {Trump}, {Vignali}, {Zamojski}, {Aldcroft},
  {Aussel}, {Bardelli}, {Bolzonella}, {Cappi}, {Caputi}, {Contini},
  {Finoguenov}, {Fruscione}, {Garilli}, {Impey}, {Iovino}, {Iwasawa},
  {Kampczyk}, {Kartaltepe}, {Kneib}, {Knobel}, {Kovac}, {Lamareille},
  {Leborgne}, {Le Brun}, {Le Fevre}, {Lilly}, {Maier}, {McCracken}, {Pello},
  {Peng}, {Perez-Montero}, {de Ravel}, {Sanders}, {Scodeggio}, {Scoville},
  {Tanaka}, {Taniguchi}, {Tasca}, {de la Torre}, {Tresse}, {Vergani}, \&
  {Zucca}}]{Brusa2010}
{Brusa}, M., {Civano}, F., {Comastri}, A., {et~al.} 2010, \apj, 716, 348

\bibitem[{{Cano-D{\'{\i}}az} {et~al.}(2012){Cano-D{\'{\i}}az}, {Maiolino},
  {Marconi}, {Netzer}, {Shemmer}, \& {Cresci}}]{CanoDiaz2012}
{Cano-D{\'{\i}}az}, M., {Maiolino}, R., {Marconi}, A., {et~al.} 2012, \aap,
  537, L8

\bibitem[{{Cappelluti} {et~al.}(2009){Cappelluti}, {Brusa}, {Hasinger},
  {Comastri}, {Zamorani}, {Finoguenov}, {Gilli}, {Puccetti}, {Miyaji},
  {Salvato}, {Vignali}, {Aldcroft}, {B{\"o}hringer}, {Brunner}, {Civano},
  {Elvis}, {Fiore}, {Fruscione}, {Griffiths}, {Guzzo}, {Iovino}, {Koekemoer},
  {Mainieri}, {Scoville}, {Shopbell}, {Silverman}, \& {Urry}}]{Cappelluti2009}
{Cappelluti}, N., {Brusa}, M., {Hasinger}, G., {et~al.} 2009, \aap, 497, 635

\bibitem[{{Cardamone} {et~al.}(2009){Cardamone}, {Schawinski}, {Sarzi},
  {Bamford}, {Bennert}, {Urry}, {Lintott}, {Keel}, {Parejko}, {Nichol},
  {Thomas}, {Andreescu}, {Murray}, {Raddick}, {Slosar}, {Szalay}, \&
  {Vandenberg}}]{Cardamone2009}
{Cardamone}, C., {Schawinski}, K., {Sarzi}, M., {et~al.} 2009, \mnras, 399,
  1191

\bibitem[{{Cardelli} {et~al.}(1989){Cardelli}, {Clayton}, \&
  {Mathis}}]{Cardelli1989}
{Cardelli}, J.~A., {Clayton}, G.~C., \& {Mathis}, J.~S. 1989, \apj, 345, 245

\bibitem[{{Carniani} {et~al.}(2015){Carniani}, {Marconi}, {Maiolino},
  {Balmaverde}, {Brusa}, {Cano-D{\'{\i}}az}, {Cicone}, {Comastri}, {Cresci},
  {Fiore}, {Feruglio}, {La Franca}, {Mainieri}, {Mannucci}, {Nagao}, {Netzer},
  {Piconcelli}, {Risaliti}, {Schneider}, \& {Shemmer}}]{Carniani2015}
{Carniani}, S., {Marconi}, A., {Maiolino}, R., {et~al.} 2015, \aap, 580, A102

\bibitem[{{Chabrier}(2003)}]{Chabrier2003}
{Chabrier}, G. 2003, \pasp, 115, 763

\bibitem[{{Chiaberge} {et~al.}(2009){Chiaberge}, {Tremblay}, {Capetti},
  {Macchetto}, {Tozzi}, \& {Sparks}}]{Chiaberge2009}
{Chiaberge}, M., {Tremblay}, G., {Capetti}, A., {et~al.} 2009, \apj, 696, 1103

\bibitem[{{Cicone} {et~al.}(2014){Cicone}, {Maiolino}, {Sturm},
  {Graci{\'a}-Carpio}, {Feruglio}, {Neri}, {Aalto}, {Davies}, {Fiore},
  {Fischer}, {Garc{\'{\i}}a-Burillo}, {Gonz{\'a}lez-Alfonso},
  {Hailey-Dunsheath}, {Piconcelli}, \& {Veilleux}}]{Cicone2014}
{Cicone}, C., {Maiolino}, R., {Sturm}, E., {et~al.} 2014, \aap, 562, A21

\bibitem[{{Cimatti} {et~al.}(2013){Cimatti}, {Brusa}, {Talia}, {Mignoli},
  {Rodighiero}, {Kurk}, {Cassata}, {Halliday}, {Renzini}, \&
  {Daddi}}]{Cimatti2013}
{Cimatti}, A., {Brusa}, M., {Talia}, M., {et~al.} 2013, \apjl, 779, L13

\bibitem[{{Civano} {et~al.}(2016){Civano}, {Marchesi}, {Comastri}, {Urry},
  {Elvis}, {Cappelluti}, {Puccetti}, {Brusa}, {Zamorani}, {Hasinger},
  {Aldcroft}, {Alexander}, {Allevato}, {Brunner}, {Capak}, {Finoguenov},
  {Fiore}, {Fruscione}, {Gilli}, {Glotfelty}, {Griffiths}, {Hao}, {Harrison},
  {Jahnke}, {Kartaltepe}, {Karim}, {LaMassa}, {Lanzuisi}, {Miyaji}, {Ranalli},
  {Salvato}, {Sargent}, {Scoville}, {Schawinski}, {Schinnerer}, {Silverman},
  {Smolcic}, {Stern}, {Toft}, {Trakhenbrot}, {Treister}, \&
  {Vignali}}]{Civano2016}
{Civano}, F., {Marchesi}, S., {Comastri}, A., {et~al.} 2016, \apj in press, arXiv:1601.00941

\bibitem[{{Contini} {et~al.}(2012){Contini}, {Garilli}, {Le F{\`e}vre},
  {Kissler-Patig}, {Amram}, {Epinat}, {Moultaka}, {Paioro}, {Queyrel}, {Tasca},
  {Tresse}, {Vergani}, {L{\'o}pez-Sanjuan}, \& {Perez-Montero}}]{Contini2012}
{Contini}, T., {Garilli}, B., {Le F{\`e}vre}, O., {et~al.} 2012, \aap, 539, A91

\bibitem[{{Cresci} {et~al.}(2009){Cresci}, {Hicks}, {Genzel}, {Schreiber},
  {Davies}, {Bouch{\'e}}, {Buschkamp}, {Genel}, {Shapiro}, {Tacconi},
  {Sommer-Larsen}, {Burkert}, {Eisenhauer}, {Gerhard}, {Lutz}, {Naab},
  {Sternberg}, {Cimatti}, {Daddi}, {Erb}, {Kurk}, {Lilly}, {Renzini},
  {Shapley}, {Steidel}, \& {Caputi}}]{Cresci2009}
{Cresci}, G., {Hicks}, E.~K.~S., {Genzel}, R., {et~al.} 2009, \apj, 697, 115

\bibitem[{{Cresci} {et~al.}(2015{\natexlab{a}}){Cresci}, {Mainieri}, {Brusa},
  {Marconi}, {Perna}, {Mannucci}, {Piconcelli}, {Maiolino}, {Feruglio},
  {Fiore}, {Bongiorno}, {Lanzuisi}, {Merloni}, {Schramm}, {Silverman}, \&
  {Civano}}]{Cresci2015}
{Cresci}, G., {Mainieri}, V., {Brusa}, M., {et~al.} 2015{\natexlab{a}}, \apj,
  799, 82

\bibitem[{{Cresci} {et~al.}(2015{\natexlab{b}}){Cresci}, {Marconi}, {Zibetti},
  {Risaliti}, {Carniani}, {Mannucci}, {Gallazzi}, {Maiolino}, {Balmaverde},
  {Brusa}, {Capetti}, {Cicone}, {Feruglio}, {Bland-Hawthorn}, {Nagao}, {Oliva},
  {Salvato}, {Sani}, {Tozzi}, {Urrutia}, \& {Venturi}}]{Cresci2015_MUSE}
{Cresci}, G., {Marconi}, A., {Zibetti}, S., {et~al.} 2015{\natexlab{b}}, \aap,
  582, A63

\bibitem[{{da Cunha} {et~al.}(2008){da Cunha}, {Charlot}, \&
  {Elbaz}}]{Dacunha2008}
{da Cunha}, E., {Charlot}, S., \& {Elbaz}, D. 2008, \mnras, 388, 1595

\bibitem[{{Davies}(2007)}]{Davies2007}
{Davies}, R.~I. 2007, \mnras, 375, 1099

\bibitem[{{Delvecchio} {et~al.}(2014){Delvecchio}, {Gruppioni}, {Pozzi},
  {Berta}, {Zamorani}, {Cimatti}, {Lutz}, {Scott}, {Vignali}, {Cresci},
  {Feltre}, {Cooray}, {Vaccari}, {Fritz}, {Le Floc'h}, {Magnelli}, {Popesso},
  {Oliver}, {Bock}, {Carollo}, {Contini}, {Le F{\'e}vre}, {Lilly}, {Mainieri},
  {Renzini}, \& {Scodeggio}}]{Delvecchio2014}
{Delvecchio}, I., {Gruppioni}, C., {Pozzi}, F., {et~al.} 2014, \mnras, 439,
  2736

\bibitem[{{Elvis} {et~al.}(2012){Elvis}, {Hao}, {Civano}, {Brusa}, {Salvato},
  {Bongiorno}, {Capak}, {Zamorani}, {Comastri}, {Jahnke}, {Lusso}, {Mainieri},
  {Trump}, {Ho}, {Aussel}, {Cappelluti}, {Cisternas}, {Frayer}, {Gilli},
  {Hasinger}, {Huchra}, {Impey}, {Koekemoer}, {Lanzuisi}, {Le Floc'h}, {Lilly},
  {Liu}, {McCarthy}, {McCracken}, {Merloni}, {Roeser}, {Sanders}, {Sargent},
  {Scoville}, {Schinnerer}, {Schiminovich}, {Silverman}, {Taniguchi},
  {Vignali}, {Urry}, {Zamojski}, \& {Zatloukal}}]{Elvis2012}
{Elvis}, M., {Hao}, H., {Civano}, F., {et~al.} 2012, \apj, 759, 6

\bibitem[{{Fabian}(2012)}]{Fabian2012}
{Fabian}, A.~C. 2012, \araa, 50, 455

\bibitem[{{Fanti} {et~al.}(1995){Fanti}, {Fanti}, {Dallacasa}, {Schilizzi},
  {Spencer}, \& {Stanghellini}}]{Fanti1995}
{Fanti}, C., {Fanti}, R., {Dallacasa}, D., {et~al.} 1995, \aap, 302, 317

\bibitem[{{Feruglio} {et~al.}(2010){Feruglio}, {Maiolino}, {Piconcelli},
  {Menci}, {Aussel}, {Lamastra}, \& {Fiore}}]{Feruglio2010}
{Feruglio}, C., {Maiolino}, R., {Piconcelli}, E., {et~al.} 2010, \aap, 518,
  L155

\bibitem[{{Floyd} {et~al.}(2008){Floyd}, {Axon}, {Baum}, {Capetti},
  {Chiaberge}, {Macchetto}, {Madrid}, {Miley}, {O'Dea}, {Perlman}, {Quillen},
  {Sparks}, \& {Tremblay}}]{Floyd2008}
{Floyd}, D.~J.~E., {Axon}, D., {Baum}, S., {et~al.} 2008, \apjs, 177, 148

\bibitem[{{F{\"o}rster Schreiber} {et~al.}(2014){F{\"o}rster Schreiber},
  {Genzel}, {Newman}, {Kurk}, {Lutz}, {Tacconi}, {Wuyts}, {Bandara}, {Burkert},
  {Buschkamp}, {Carollo}, {Cresci}, {Daddi}, {Davies}, {Eisenhauer}, {Hicks},
  {Lang}, {Lilly}, {Mainieri}, {Mancini}, {Naab}, {Peng}, {Renzini}, {Rosario},
  {Shapiro Griffin}, {Shapley}, {Sternberg}, {Tacchella}, {Vergani},
  {Wisnioski}, {Wuyts}, \& {Zamorani}}]{FS2014}
{F{\"o}rster Schreiber}, N.~M., {Genzel}, R., {Newman}, S.~F., {et~al.} 2014,
  \apj, 787, 38

\bibitem[{{Gaskell} \& {Goosmann}(2013)}]{Gaskell2013}
{Gaskell}, C.~M. \& {Goosmann}, R.~W. 2013, \apj, 769, 30

\bibitem[{{Geach} {et~al.}(2014){Geach}, {Hickox}, {Diamond-Stanic}, {Krips},
  {Rudnick}, {Tremonti}, {Sell}, {Coil}, \& {Moustakas}}]{Geach2014}
{Geach}, J.~E., {Hickox}, R.~C., {Diamond-Stanic}, A.~M., {et~al.} 2014, \nat,
  516, 68

\bibitem[{{Gelderman} \& {Whittle}(1994)}]{Gelderman1994}
{Gelderman}, R. \& {Whittle}, M. 1994, \apjs, 91, 491

\bibitem[{{Genzel} {et~al.}(2014){Genzel}, {F{\"o}rster Schreiber}, {Rosario},
  {Lang}, {Lutz}, {Wisnioski}, {Wuyts}, {Wuyts}, {Bandara}, {Bender}, {Berta},
  {Kurk}, {Mendel}, {Tacconi}, {Wilman}, {Beifiori}, {Brammer}, {Burkert},
  {Buschkamp}, {Chan}, {Carollo}, {Davies}, {Eisenhauer}, {Fabricius},
  {Fossati}, {Kriek}, {Kulkarni}, {Lilly}, {Mancini}, {Momcheva}, {Naab},
  {Nelson}, {Renzini}, {Saglia}, {Sharples}, {Sternberg}, {Tacchella}, \& {van
  Dokkum}}]{Genzel2014}
{Genzel}, R., {F{\"o}rster Schreiber}, N.~M., {Rosario}, D., {et~al.} 2014,
  \apj, 796, 7

\bibitem[{{Hao} {et~al.}(2014){Hao}, {Elvis}, {Civano}, {Zamorani}, {Ho},
  {Comastri}, {Brusa}, {Bongiorno}, {Merloni}, {Trump}, {Salvato}, {Impey},
  {Koekemoer}, {Lanzuisi}, {Celotti}, {Jahnke}, {Vignali}, {Silverman}, {Urry},
  {Schawinski}, \& {Capak}}]{Hao2014}
{Hao}, H., {Elvis}, M., {Civano}, F., {et~al.} 2014, \mnras, 438, 1288

\bibitem[{{Harrison} {et~al.}(2016){Harrison}, {Alexander}, {Mullaney},
  {Stott}, {Swinbank}, {Arumugam}, {Bauer}, {Bower}, {Bunker}, \&
  {Sharples}}]{Harrison2016}
{Harrison}, C.~M., {Alexander}, D.~M., {Mullaney}, J.~R., {et~al.} 2016,
  \mnras, 456, 1195

\bibitem[{{Harrison} {et~al.}(2014){Harrison}, {Alexander}, {Mullaney}, \&
  {Swinbank}}]{Harrison2014}
{Harrison}, C.~M., {Alexander}, D.~M., {Mullaney}, J.~R., \& {Swinbank}, A.~M.
  2014, \mnras, 441, 3306

\bibitem[{{Harrison} {et~al.}(2012){Harrison}, {Alexander}, {Swinbank},
  {Smail}, {Alaghband-Zadeh}, {Bauer}, {Chapman}, {Del Moro}, {Hickox},
  {Ivison}, {Men{\'e}ndez-Delmestre}, {Mullaney}, \& {Nesvadba}}]{Harrison2012}
{Harrison}, C.~M., {Alexander}, D.~M., {Swinbank}, A.~M., {et~al.} 2012,
  \mnras, 426, 1073

\bibitem[{{Hasinger} {et~al.}(2007){Hasinger}, {Cappelluti}, {Brunner},
  {Brusa}, {Comastri}, {Elvis}, {Finoguenov}, {Fiore}, {Franceschini}, {Gilli},
  {Griffiths}, {Lehmann}, {Mainieri}, {Matt}, {Matute}, {Miyaji}, {Molendi},
  {Paltani}, {Sanders}, {Scoville}, {Tresse}, {Urry}, {Vettolani}, \&
  {Zamorani}}]{Hasinger2007}
{Hasinger}, G., {Cappelluti}, N., {Brunner}, H., {et~al.} 2007, \apjs, 172, 29

\bibitem[{{Hayano} {et~al.}(2010){Hayano}, {Takami}, {Oya}, {Hattori}, {Saito},
  {Watanabe}, {Guyon}, {Minowa}, {Egner}, {Ito}, {Garrel}, {Colley}, {Golota},
  \& {Iye}}]{Hayano2010}
{Hayano}, Y., {Takami}, H., {Oya}, S., {et~al.} 2010, in \procspie, Vol. 7736,
  Adaptive Optics Systems II, 77360N

\bibitem[{{Heckman} {et~al.}(1990){Heckman}, {Armus}, \& {Miley}}]{Heckman1990}
{Heckman}, T.~M., {Armus}, L., \& {Miley}, G.~K. 1990, \apjs, 74, 833

\bibitem[{{Hopkins} {et~al.}(2006){Hopkins}, {Hernquist}, {Cox}, {Di Matteo},
  {Robertson}, \& {Springel}}]{Hopkins2006a}
{Hopkins}, P.~F., {Hernquist}, L., {Cox}, T.~J., {et~al.} 2006, \apjs, 163, 1

\bibitem[{{Hopkins} {et~al.}(2008){Hopkins}, {Hernquist}, {Cox}, \& {Kere{\v
  s}}}]{Hopkins2008}
{Hopkins}, P.~F., {Hernquist}, L., {Cox}, T.~J., \& {Kere{\v s}}, D. 2008,
  \apjs, 175, 356

\bibitem[{{Humphrey} {et~al.}(2008){Humphrey}, {Villar-Mart{\'{\i}}n},
  {Vernet}, {Fosbury}, {di Serego Alighieri}, \& {Binette}}]{Humphrey2008}
{Humphrey}, A., {Villar-Mart{\'{\i}}n}, M., {Vernet}, J., {et~al.} 2008,
  \mnras, 383, 11

\bibitem[{{Jones} {et~al.}(2015){Jones}, {Martin}, \& {Cooper}}]{Jones2015}
{Jones}, T., {Martin}, C., \& {Cooper}, M.~C. 2015, \apj, 813, 126

\bibitem[{{Kewley} {et~al.}(2013){Kewley}, {Maier}, {Yabe}, {Ohta}, {Akiyama},
  {Dopita}, \& {Yuan}}]{Kewley2013}
{Kewley}, L.~J., {Maier}, C., {Yabe}, K., {et~al.} 2013, \apjl, 774, L10

\bibitem[{{Kim} {et~al.}(2013){Kim}, {Ho}, {Lonsdale}, {Lacy}, {Blain}, \&
  {Kimball}}]{Kim2013}
{Kim}, M., {Ho}, L.~C., {Lonsdale}, C.~J., {et~al.} 2013, \apjl, 768, L9

\bibitem[{{King}(2005)}]{King2005}
{King}, A. 2005, \apjl, 635, L121

\bibitem[{{King} \& {Pounds}(2015)}]{King2015}
{King}, A. \& {Pounds}, K. 2015, \araa, 53, 115

\bibitem[{{King} {et~al.}(2011){King}, {Zubovas}, \& {Power}}]{King2011}
{King}, A.~R., {Zubovas}, K., \& {Power}, C. 2011, \mnras, 415, L6

\bibitem[{{Kobayashi} {et~al.}(2000){Kobayashi}, {Tokunaga}, {Terada}, {Goto},
  {Weber}, {Potter}, {Onaka}, {Ching}, {Young}, {Fletcher}, {Neil},
  {Robertson}, {Cook}, {Imanishi}, \& {Warren}}]{Kobayashi2000}
{Kobayashi}, N., {Tokunaga}, A.~T., {Terada}, H., {et~al.} 2000, in \procspie,
  Vol. 4008, Optical and IR Telescope Instrumentation and Detectors, ed.
  M.~{Iye} \& A.~F. {Moorwood}, 1056--1066

\bibitem[{{Koekemoer} {et~al.}(2007){Koekemoer}, {Aussel}, {Calzetti}, {Capak},
  {Giavalisco}, {Kneib}, {Leauthaud}, {Le F{\`e}vre}, {McCracken}, {Massey},
  {Mobasher}, {Rhodes}, {Scoville}, \& {Shopbell}}]{Koekemoer2007}
{Koekemoer}, A.~M., {Aussel}, H., {Calzetti}, D., {et~al.} 2007, \apjs, 172,
  196

\bibitem[{{Kormendy} \& {Ho}(2013)}]{Kormendy2013}
{Kormendy}, J. \& {Ho}, L.~C. 2013, \araa, 51, 511

\bibitem[{{Lamareille} {et~al.}(2009){Lamareille}, {Brinchmann}, {Contini},
  {Walcher}, {Charlot}, {P{\'e}rez-Montero}, {Zamorani}, {Pozzetti},
  {Bolzonella}, {Garilli}, {Paltani}, {Bongiorno}, {Le F{\`e}vre}, {Bottini},
  {Le Brun}, {Maccagni}, {Scaramella}, {Scodeggio}, {Tresse}, {Vettolani},
  {Zanichelli}, {Adami}, {Arnouts}, {Bardelli}, {Cappi}, {Ciliegi}, {Foucaud},
  {Franzetti}, {Gavignaud}, {Guzzo}, {Ilbert}, {Iovino}, {McCracken}, {Marano},
  {Marinoni}, {Mazure}, {Meneux}, {Merighi}, {Pell{\`o}}, {Pollo}, {Radovich},
  {Vergani}, {Zucca}, {Romano}, {Grado}, \& {Limatola}}]{Lamareille2009}
{Lamareille}, F., {Brinchmann}, J., {Contini}, T., {et~al.} 2009, \aap, 495, 53

\bibitem[{{Lanz} {et~al.}(2015){Lanz}, {Ogle}, {Evans}, {Appleton}, {Guillard},
  \& {Emonts}}]{Lanz2015}
{Lanz}, L., {Ogle}, P.~M., {Evans}, D., {et~al.} 2015, \apj, 801, 17

\bibitem[{{Lanzuisi} {et~al.}(2015){Lanzuisi}, {Ranalli}, {Georgantopoulos},
  {Georgakakis}, {Delvecchio}, {Akylas}, {Berta}, {Bongiorno}, {Brusa},
  {Cappelluti}, {Civano}, {Comastri}, {Gilli}, {Gruppioni}, {Hasinger},
  {Iwasawa}, {Koekemoer}, {Lusso}, {Marchesi}, {Mainieri}, {Merloni},
  {Mignoli}, {Piconcelli}, {Pozzi}, {Rosario}, {Salvato}, {Silverman},
  {Trakhtenbrot}, {Vignali}, \& {Zamorani}}]{Lanzuisi2015}
{Lanzuisi}, G., {Ranalli}, P., {Georgantopoulos}, I., {et~al.} 2015, \aap, 573,
  A137

\bibitem[{{Lilly} {et~al.}(2009){Lilly}, {LeBrun}, {Maier}, {Mainieri},
  {Mignoli}, {Scodeggio}, {Zamorani}, {Carollo}, {Contini}, {Kneib},
  {LeF{\`e}vre}, {Renzini}, {Bardelli}, {Bolzonella}, {Bongiorno}, {Caputi},
  {Coppa}, {Cucciati}, {de la Torre}, {de Ravel}, {Franzetti}, {Garilli},
  {Iovino}, {Kampczyk}, {Kovac}, {Knobel}, {Lamareille}, {LeBorgne}, {Pello},
  {Peng}, {P{\'e}rez-Montero}, {Ricciardelli}, {Silverman}, {Tanaka}, {Tasca},
  {Tresse}, {Vergani}, {Zucca}, {Ilbert}, {Salvato}, {Oesch}, {Abbas},
  {Bottini}, {Capak}, {Cappi}, {Cassata}, {Cimatti}, {Elvis}, {Fumana},
  {Guzzo}, {Hasinger}, {Koekemoer}, {Leauthaud}, {Maccagni}, {Marinoni},
  {McCracken}, {Memeo}, {Meneux}, {Porciani}, {Pozzetti}, {Sanders},
  {Scaramella}, {Scarlata}, {Scoville}, {Shopbell}, \& {Taniguchi}}]{Lilly2009}
{Lilly}, S.~J., {LeBrun}, V., {Maier}, C., {et~al.} 2009, \apjs, 184, 218

\bibitem[{{Liu} {et~al.}(2013){Liu}, {Zakamska}, {Greene}, {Nesvadba}, \&
  {Liu}}]{Liu2013}
{Liu}, G., {Zakamska}, N.~L., {Greene}, J.~E., {Nesvadba}, N.~P.~H., \& {Liu},
  X. 2013, \mnras, 436, 2576

\bibitem[{{Lusso} {et~al.}(2012){Lusso}, {Comastri}, {Simmons}, {Mignoli},
  {Zamorani}, {Vignali}, {Brusa}, {Shankar}, {Lutz}, {Trump}, {Maiolino},
  {Gilli}, {Bolzonella}, {Puccetti}, {Salvato}, {Impey}, {Civano}, {Elvis},
  {Mainieri}, {Silverman}, {Koekemoer}, {Bongiorno}, {Merloni}, {Berta}, {Le
  Floc'h}, {Magnelli}, {Pozzi}, \& {Riguccini}}]{Lusso2012}
{Lusso}, E., {Comastri}, A., {Simmons}, B.~D., {et~al.} 2012, \mnras, 425, 623

\bibitem[{{Magliocchetti} {et~al.}(2014){Magliocchetti}, {Lutz}, {Rosario},
  {Berta}, {Le Floc'h}, {Magnelli}, {Pozzi}, {Riguccini}, \&
  {Santini}}]{Magliocchetti2014}
{Magliocchetti}, M., {Lutz}, D., {Rosario}, D., {et~al.} 2014, \mnras, 442, 682

\bibitem[{{Magorrian} {et~al.}(1998){Magorrian}, {Tremaine}, {Richstone},
  {Bender}, {Bower}, {Dressler}, {Faber}, {Gebhardt}, {Green}, {Grillmair},
  {Kormendy}, \& {Lauer}}]{Magorrian1998}
{Magorrian}, J., {Tremaine}, S., {Richstone}, D., {et~al.} 1998, \aj, 115, 2285

\bibitem[{{Mahony} {et~al.}(2016){Mahony}, {Oonk}, {Morganti}, {Tadhunter},
  {Bessiere}, {Short}, {Emonts}, \& {Oosterloo}}]{Mahony2016}
{Mahony}, E.~K., {Oonk}, J.~B.~R., {Morganti}, R., {et~al.} 2016, \mnras, 455,
  2453

\bibitem[{{Mainieri} {et~al.}(2011){Mainieri}, {Bongiorno}, {Merloni}, {Aller},
  {Carollo}, {Iwasawa}, {Koekemoer}, {Mignoli}, {Silverman}, {Bolzonella},
  {Brusa}, {Comastri}, {Gilli}, {Halliday}, {Ilbert}, {Lusso}, {Salvato},
  {Vignali}, {Zamorani}, {Contini}, {Kneib}, {Le F{\`e}vre}, {Lilly},
  {Renzini}, {Scodeggio}, {Balestra}, {Bardelli}, {Caputi}, {Coppa},
  {Cucciati}, {de la Torre}, {de Ravel}, {Franzetti}, {Garilli}, {Iovino},
  {Kampczyk}, {Knobel}, {Kova{\v c}}, {Lamareille}, {Le Borgne}, {Le Brun},
  {Maier}, {Nair}, {Pello}, {Peng}, {Perez Montero}, {Pozzetti},
  {Ricciardelli}, {Tanaka}, {Tasca}, {Tresse}, {Vergani}, {Zucca}, {Aussel},
  {Capak}, {Cappelluti}, {Elvis}, {Fiore}, {Hasinger}, {Impey}, {Le Floc'h},
  {Scoville}, {Taniguchi}, \& {Trump}}]{Mainieri2011}
{Mainieri}, V., {Bongiorno}, A., {Merloni}, A., {et~al.} 2011, \aap, 535, A80

\bibitem[{{Maiolino} {et~al.}(2012){Maiolino}, {Gallerani}, {Neri}, {Cicone},
  {Ferrara}, {Genzel}, {Lutz}, {Sturm}, {Tacconi}, {Walter}, {Feruglio},
  {Fiore}, \& {Piconcelli}}]{Maiolino2012}
{Maiolino}, R., {Gallerani}, S., {Neri}, R., {et~al.} 2012, \mnras, 425, L66

\bibitem[{{Mancini} {et~al.}(2011){Mancini}, {F{\"o}rster Schreiber},
  {Renzini}, {Cresci}, {Hicks}, {Peng}, {Vergani}, {Lilly}, {Carollo},
  {Pozzetti}, {Zamorani}, {Daddi}, {Genzel}, {Maraston}, {McCracken},
  {Tacconi}, {Bouch{\'e}}, {Davies}, {Oesch}, {Shapiro}, {Mainieri}, {Lutz},
  {Mignoli}, \& {Sternberg}}]{Mancini2011}
{Mancini}, C., {F{\"o}rster Schreiber}, N.~M., {Renzini}, A., {et~al.} 2011,
  \apj, 743, 86

\bibitem[{{Marconi} \& {Hunt}(2003)}]{Marconi2003}
{Marconi}, A. \& {Hunt}, L.~K. 2003, \apjl, 589, L21

\bibitem[{{Marziani} {et~al.}(2016){Marziani}, {Sulentic}, {Stirpe}, {Dultzin},
  {Del Olmo}, \& {Mart{\'{\i}}nez-Carballo}}]{Marziani2016}
{Marziani}, P., {Sulentic}, J.~W., {Stirpe}, G.~M., {et~al.} 2016, \apss, 361,
  3

\bibitem[{{McCracken} {et~al.}(2012){McCracken}, {Milvang-Jensen}, {Dunlop},
  {Franx}, {Fynbo}, {Le F{\`e}vre}, {Holt}, {Caputi}, {Goranova}, {Buitrago},
  {Emerson}, {Freudling}, {Hudelot}, {L{\'o}pez-Sanjuan}, {Magnard}, {Mellier},
  {M{\o}ller}, {Nilsson}, {Sutherland}, {Tasca}, \& {Zabl}}]{McCracken2012}
{McCracken}, H.~J., {Milvang-Jensen}, B., {Dunlop}, J., {et~al.} 2012, \aap,
  544, A156

\bibitem[{{Merloni}(2016)}]{Merloni2016}
{Merloni}, A. 2016, in Lecture Notes in Physics, Berlin Springer Verlag, Vol.
  905, Lecture Notes in Physics, Berlin Springer Verlag, ed. F.~{Haardt},
  V.~{Gorini}, U.~{Moschella}, A.~{Treves}, \& M.~{Colpi}, 101

\bibitem[{{Merloni} {et~al.}(2014){Merloni}, {Bongiorno}, {Brusa}, {Iwasawa},
  {Mainieri}, {Magnelli}, {Salvato}, {Berta}, {Cappelluti}, {Comastri},
  {Fiore}, {Gilli}, {Koekemoer}, {Le Floc'h}, {Lusso}, {Lutz}, {Miyaji},
  {Pozzi}, {Riguccini}, {Rosario}, {Silverman}, {Symeonidis}, {Treister},
  {Vignali}, \& {Zamorani}}]{Merloni2014}
{Merloni}, A., {Bongiorno}, A., {Brusa}, M., {et~al.} 2014, \mnras, 437, 3550

\bibitem[{{Nakajima} \& {Ouchi}(2014)}]{Nakajima2014}
{Nakajima}, K. \& {Ouchi}, M. 2014, \mnras, 442, 900

\bibitem[{{Orienti}(2015)}]{Orienti2015}
{Orienti}, M. 2015, Astr. Nat. in press, arXiv:1511.00436

\bibitem[{{Perna} {et~al.}(2015{\natexlab{a}}){Perna}, {Brusa}, {Cresci},
  {Comastri}, {Lanzuisi}, {Lusso}, {Marconi}, {Salvato}, {Zamorani},
  {Bongiorno}, {Mainieri}, {Maiolino}, \& {Mignoli}}]{Perna2015}
{Perna}, M., {Brusa}, M., {Cresci}, G., {et~al.} 2015{\natexlab{a}}, \aap, 574,
  A82

\bibitem[{{Perna} {et~al.}(2015{\natexlab{b}}){Perna}, {Brusa}, {Salvato},
  {Cresci}, {Lanzuisi}, {Berta}, {Delvecchio}, {Fiore}, {Lutz}, {Le Floc'h},
  {Mainieri}, \& {Riguccini}}]{Perna2015_miro}
{Perna}, M., {Brusa}, M., {Salvato}, M., {et~al.} 2015{\natexlab{b}}, \aap,
  583, A72

\bibitem[{{Rodighiero} {et~al.}(2011){Rodighiero}, {Daddi}, {Baronchelli},
  {Cimatti}, {Renzini}, {Aussel}, {Popesso}, {Lutz}, {Andreani}, {Berta},
  {Cava}, {Elbaz}, {Feltre}, {Fontana}, {F{\"o}rster Schreiber},
  {Franceschini}, {Genzel}, {Grazian}, {Gruppioni}, {Ilbert}, {Le Floch},
  {Magdis}, {Magliocchetti}, {Magnelli}, {Maiolino}, {McCracken}, {Nordon},
  {Poglitsch}, {Santini}, {Pozzi}, {Riguccini}, {Tacconi}, {Wuyts}, \&
  {Zamorani}}]{Rodighiero2011}
{Rodighiero}, G., {Daddi}, E., {Baronchelli}, I., {et~al.} 2011, \apjl, 739,
  L40

\bibitem[{{Rodr{\'{\i}}guez-Zaur{\'{\i}}n}
  {et~al.}(2013){Rodr{\'{\i}}guez-Zaur{\'{\i}}n}, {Tadhunter}, {Rose}, \&
  {Holt}}]{Zaurin2013}
{Rodr{\'{\i}}guez-Zaur{\'{\i}}n}, J., {Tadhunter}, C.~N., {Rose}, M., \&
  {Holt}, J. 2013, \mnras, 432, 138

\bibitem[{{Rupke} \& {Veilleux}(2011)}]{Rupke2011}
{Rupke}, D.~S.~N. \& {Veilleux}, S. 2011, \apjl, 729, L27

\bibitem[{{Rupke} \& {Veilleux}(2013)}]{Rupke2013}
{Rupke}, D.~S.~N. \& {Veilleux}, S. 2013, \apjl, 775, L15

\bibitem[{{Schinnerer} {et~al.}(2010){Schinnerer}, {Sargent}, {Bondi}, {Smol{\v
  c}i{\'c}}, {Datta}, {Carilli}, {Bertoldi}, {Blain}, {Ciliegi}, {Koekemoer},
  \& {Scoville}}]{Schinnerer2010}
{Schinnerer}, E., {Sargent}, M.~T., {Bondi}, M., {et~al.} 2010, \apjs, 188, 384

\bibitem[{{Scoville} {et~al.}(2007{\natexlab{a}}){Scoville}, {Aussel},
  {Benson}, {Blain}, {Calzetti}, {Capak}, {Ellis}, {El-Zant}, {Finoguenov},
  {Giavalisco}, {Guzzo}, {Hasinger}, {Koda}, {Le F{\`e}vre}, {Massey},
  {McCracken}, {Mobasher}, {Renzini}, {Rhodes}, {Salvato}, {Sanders}, {Sasaki},
  {Schinnerer}, {Sheth}, {Shopbell}, {Taniguchi}, {Taylor}, \&
  {Thompson}}]{Scoville2007_struct}
{Scoville}, N., {Aussel}, H., {Benson}, A., {et~al.} 2007{\natexlab{a}}, \apjs,
  172, 150

\bibitem[{{Scoville} {et~al.}(2007{\natexlab{b}}){Scoville}, {Aussel}, {Brusa},
  {Capak}, {Carollo}, {Elvis}, {Giavalisco}, {Guzzo}, {Hasinger}, {Impey},
  {Kneib}, {LeFevre}, {Lilly}, {Mobasher}, {Renzini}, {Rich}, {Sanders},
  {Schinnerer}, {Schminovich}, {Shopbell}, {Taniguchi}, \&
  {Tyson}}]{Scoville2007}
{Scoville}, N., {Aussel}, H., {Brusa}, M., {et~al.} 2007{\natexlab{b}}, \apjs,
  172, 1

\bibitem[{{Stern} {et~al.}(2015){Stern}, {Faucher-Giguere}, {Zakamska}, \&
  {Hennawi}}]{Stern2015_out}
{Stern}, J., {Faucher-Giguere}, C.-A., {Zakamska}, N.~L., \& {Hennawi}, J.~F.
  2015, arXiv:1510.07690

\bibitem[{{Sturm} {et~al.}(2011){Sturm}, {Gonz{\'a}lez-Alfonso}, {Veilleux},
  {Fischer}, {Graci{\'a}-Carpio}, {Hailey-Dunsheath}, {Contursi}, {Poglitsch},
  {Sternberg}, {Davies}, {Genzel}, {Lutz}, {Tacconi}, {Verma}, {Maiolino}, \&
  {de Jong}}]{Sturm2011}
{Sturm}, E., {Gonz{\'a}lez-Alfonso}, E., {Veilleux}, S., {et~al.} 2011, \apjl,
  733, L16

\bibitem[{{Urrutia} {et~al.}(2008){Urrutia}, {Lacy}, \& {Becker}}]{Urrutia2008}
{Urrutia}, T., {Lacy}, M., \& {Becker}, R.~H. 2008, \apj, 674, 80

\bibitem[{{Vanden Berk} {et~al.}(2001){Vanden Berk}, {Richards}, {Bauer},
  {Strauss}, {Schneider}, {Heckman}, {York}, {Hall}, {Fan}, {Knapp},
  {Anderson}, {Annis}, {Bahcall}, {Bernardi}, {Briggs}, \&
  al.}]{Vandenberk2001}
{Vanden Berk}, D.~E., {Richards}, G.~T., {Bauer}, A., {et~al.} 2001, \aj, 122,
  549

\bibitem[{{Veilleux} {et~al.}(2005){Veilleux}, {Cecil}, \&
  {Bland-Hawthorn}}]{Veilleux2005}
{Veilleux}, S., {Cecil}, G., \& {Bland-Hawthorn}, J. 2005, \araa, 43, 769

\bibitem[{{Villar-Mart{\'{\i}}n} {et~al.}(2011){Villar-Mart{\'{\i}}n},
  {Humphrey}, {Delgado}, {Colina}, \& {Arribas}}]{Villar2011a}
{Villar-Mart{\'{\i}}n}, M., {Humphrey}, A., {Delgado}, R.~G., {Colina}, L., \&
  {Arribas}, S. 2011, \mnras, 418, 2032

\bibitem[{{Volonteri} {et~al.}(2015){Volonteri}, {Capelo}, {Netzer},
  {Bellovary}, {Dotti}, \& {Governato}}]{Volonteri2015}
{Volonteri}, M., {Capelo}, P.~R., {Netzer}, H., {et~al.} 2015, \mnras, 449,
  1470

\bibitem[{{Wagner} {et~al.}(2013){Wagner}, {Umemura}, \&
  {Bicknell}}]{Wagner2013}
{Wagner}, A.~Y., {Umemura}, M., \& {Bicknell}, G.~V. 2013, \apjl, 763, L18

\bibitem[{{Whitaker} {et~al.}(2012){Whitaker}, {van Dokkum}, {Brammer}, \&
  {Franx}}]{Whitaker2012}
{Whitaker}, K.~E., {van Dokkum}, P.~G., {Brammer}, G., \& {Franx}, M. 2012,
  \apjl, 754, L29

\bibitem[{{Zakamska} \& {Greene}(2014)}]{Zakamska2014}
{Zakamska}, N.~L. \& {Greene}, J.~E. 2014, \mnras, 442, 784

\bibitem[{{Zakamska} {et~al.}(2015){Zakamska}, {Hamann}, {P{\^a}ris}, {Brandt},
  {Greene}, {Strauss}, {Villforth}, {Wylezalek}, {Alexandroff}, \&
  {Ross}}]{Zakamska2016_xshooter}
{Zakamska}, N.~L., {Hamann}, F., {P{\^a}ris}, I., {et~al.} 2015, arXiv:1512.02642

\bibitem[{{Zakamska} {et~al.}(2016){Zakamska}, {Lampayan}, {Petric}, {Dicken},
  {Greene}, {Heckman}, {Hickox}, {Ho}, {Krolik}, {Nesvadba}, {Strauss},
  {Geach}, {Oguri}, \& {Strateva}}]{Zakamska2016_radio}
{Zakamska}, N.~L., {Lampayan}, K., {Petric}, A., {et~al.} 2016, \mnras, 455,
  4191

\bibitem[{{Zakamska} {et~al.}(2003){Zakamska}, {Strauss}, {Krolik}, {Collinge},
  {Hall}, {Hao}, {Heckman}, {Ivezi{\'c}}, {Richards}, {Schlegel}, {Schneider},
  {Strateva}, {Vanden Berk}, {Anderson}, \& {Brinkmann}}]{Zakamska2003}
{Zakamska}, N.~L., {Strauss}, M.~A., {Krolik}, J.~H., {et~al.} 2003, \aj, 126,
  2125

\bibitem[{{Zamanov} {et~al.}(2002){Zamanov}, {Marziani}, {Sulentic}, {Calvani},
  {Dultzin-Hacyan}, \& {Bachev}}]{Zamanov2002}
{Zamanov}, R., {Marziani}, P., {Sulentic}, J.~W., {et~al.} 2002, \apjl, 576, L9

\bibitem[{{Zhang} {et~al.}(2011){Zhang}, {Dong}, {Wang}, \&
  {Gaskell}}]{Zhang2011}
{Zhang}, K., {Dong}, X.-B., {Wang}, T.-G., \& {Gaskell}, C.~M. 2011, \apj, 737,
  71

\bibitem[{{Zubovas} \& {King}(2012)}]{Zubovas2012}
{Zubovas}, K. \& {King}, A. 2012, \apjl, 745, L34

\end{thebibliography}

%\end{document}

\end{document}